\documentclass[superscriptaddress,nobibnotes,amsmath,amssymb,notitlepage,twocolumn,prl,longbibliography]{revtex4-1}

\usepackage{setspace} 
\usepackage{graphicx} \graphicspath{{./}} 
\usepackage{amsmath}
\usepackage{color}
\usepackage{amsmath}
\usepackage{amssymb}
\usepackage{verbatim}
\usepackage{latexsym}
\usepackage{enumerate} 
\usepackage{bm} 
\usepackage{ulem} 

\begin{document}

\title{
Negative and positive anisotropic thermal expansion in 2D fullerene networks
}



\newcommand{\TCM}{Theory of Condensed Matter Group, Cavendish Laboratory, University of Cambridge, J.\,J.\,Thomson Avenue, Cambridge CB3 0HE, UK}
\newcommand{\HarvardFAS}{Harvard University, Cambridge, Massachusetts 02138, USA}
\newcommand{\Pet}{Peterhouse, University of Cambridge, Trumpington Street, Cambridge CB2 1RD, UK}


\author{Armaan Shaikh}
\affiliation{Homerton College, University of Cambridge, Hills Road, Cambridge, CB2 8PH, UK}
\affiliation{Department of Chemistry and Chemical Biology, \HarvardFAS}

\author{Jiaqi Wu}
\affiliation{\Pet}

\author{Bo Peng}
\email{bp432@cam.ac.uk}
\affiliation{\TCM}

\date{\today}

\begin{abstract}
We find a design principle for tailoring thermal expansion properties in nearly-spherical molecular networks. Using 2D fullerene networks as a representative system, we realize positive thermal expansion along intermolecular [2\,+\,2] cycloaddition bonds and negative thermal expansion along intermolecular C$-$C single bonds by varying the structural frameworks of molecules. The microscopic mechanism originates from a combination of the framework's geometric flexibility and its transverse vibrational characteristics. Based on this insight, we find molecular networks beyond C$_{60}$ with tunable thermal expansion. These findings shed light on the fundamental mechanisms governing thermal expansion in molecular networks towards rational materials design.
\end{abstract}

\maketitle


Thermal expansion is a fundamental property of materials that indicates increases in length, area, or volume upon heating, which is important in applications such as construction\,\cite{Johnson1944}, seismographs\,\cite{Mohn1999}, and aerospace design\,\cite{Strock1992,Toropova2015}. Positive thermal expansion occurs as a result of an anharmonic potential energy surface, where interatomic distance increases with increasing temperature\,\cite{Barrera2005,Liu2017c}. Counterintuitively, some materials exhibit negative thermal expansion, where increasing temperature leads to a contraction along certain crystallographic directions\,\cite{Baughman1993,Mary1996,Das2010,Fortes2011,Takenaka2012,Liu2018b,Ritz2018,Shi2021,Li2022a}. Such behaviors are attributed to flexible crystalline networks\,\cite{Goodwin2008,Dove2016}, rigid unit modes\,\cite{Pryde1996,Sleight1998,Tucker2005,Tan2024}, and transverse displacements of bridging atoms\,\cite{Hancock2004,Goodwin2005} or membranes\,\cite{Huang2016c,Koocher2021}. However, a general design principle for developing materials with negative thermal expansion is still lacking.

\begin{figure}
\centering
\includegraphics[width=0.88\linewidth]{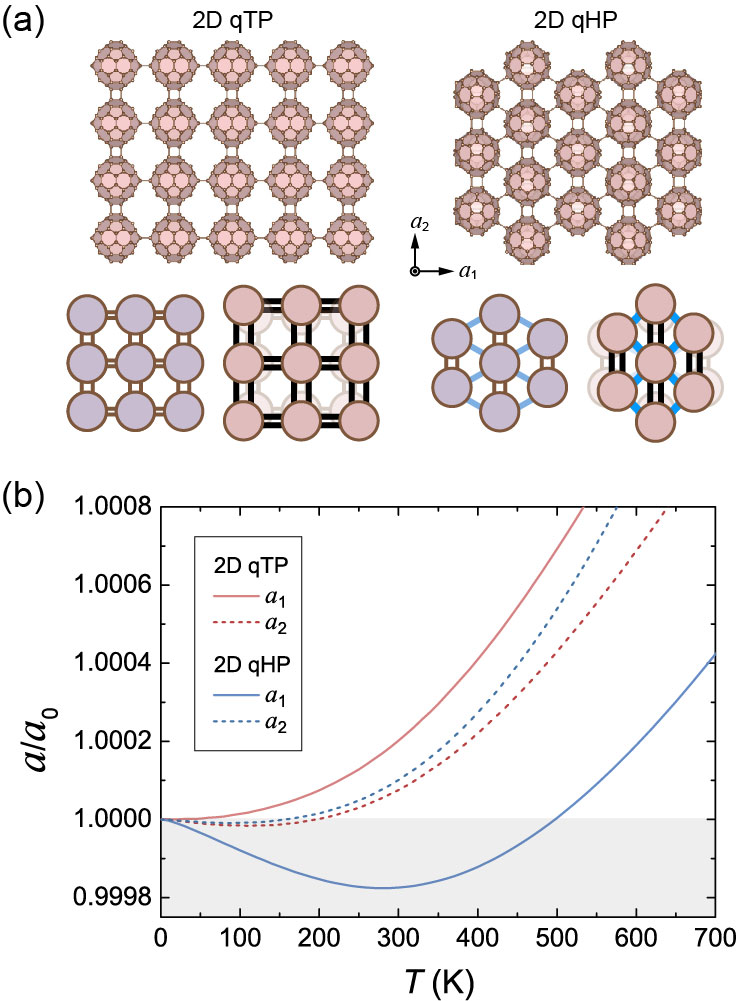}
\caption{
(a) Crystal structures and (b) thermal expansion of monolayer qTP and qHP C$_{60}$ networks. The schematics in (a) show the structural changes with increased temperature.
}
\label{crystals} 
\end{figure}

Recent synthesis of monolayer C$_{60}$ networks\,\cite{Hou2022} provides new avenues for designing materials with tunable thermal expansion. These networks exhibit diverse crystalline frameworks\,\cite{Hou2022,Peng2022c,Peng2023,Jones2023,Shearsby2025,Kayley2025,Peng2025a,Peng2025c,Meirzadeh2023} with nearly-spherical, stable units\,\cite{Tromer2022,Ribeiro2022} beyond rigid unit modes, as well as various intermolecular bridge bonds\,\cite{Zhao2024a} with tunable transverse displacements. Compared to thermal expansion in C$_{60}$ molecules and solids\,\cite{Nagel1999,Arvanitidis2003,Kwon2004,Brown2006}, the thermal behavior of C$_{60}$ monolayers has yet to be understood. In 2D form, the rotational degree of freedom of C$_{60}$ leads to different types of intermolecular bonds in varied crystalline networks [Fig.\,\ref{crystals}(a)]. In this context, it would be insightful to study whether thermal expansion in molecular networks can be controlled by intermolecular bridge bonds.




\begin{figure*}
\centering
\includegraphics[width=\linewidth]{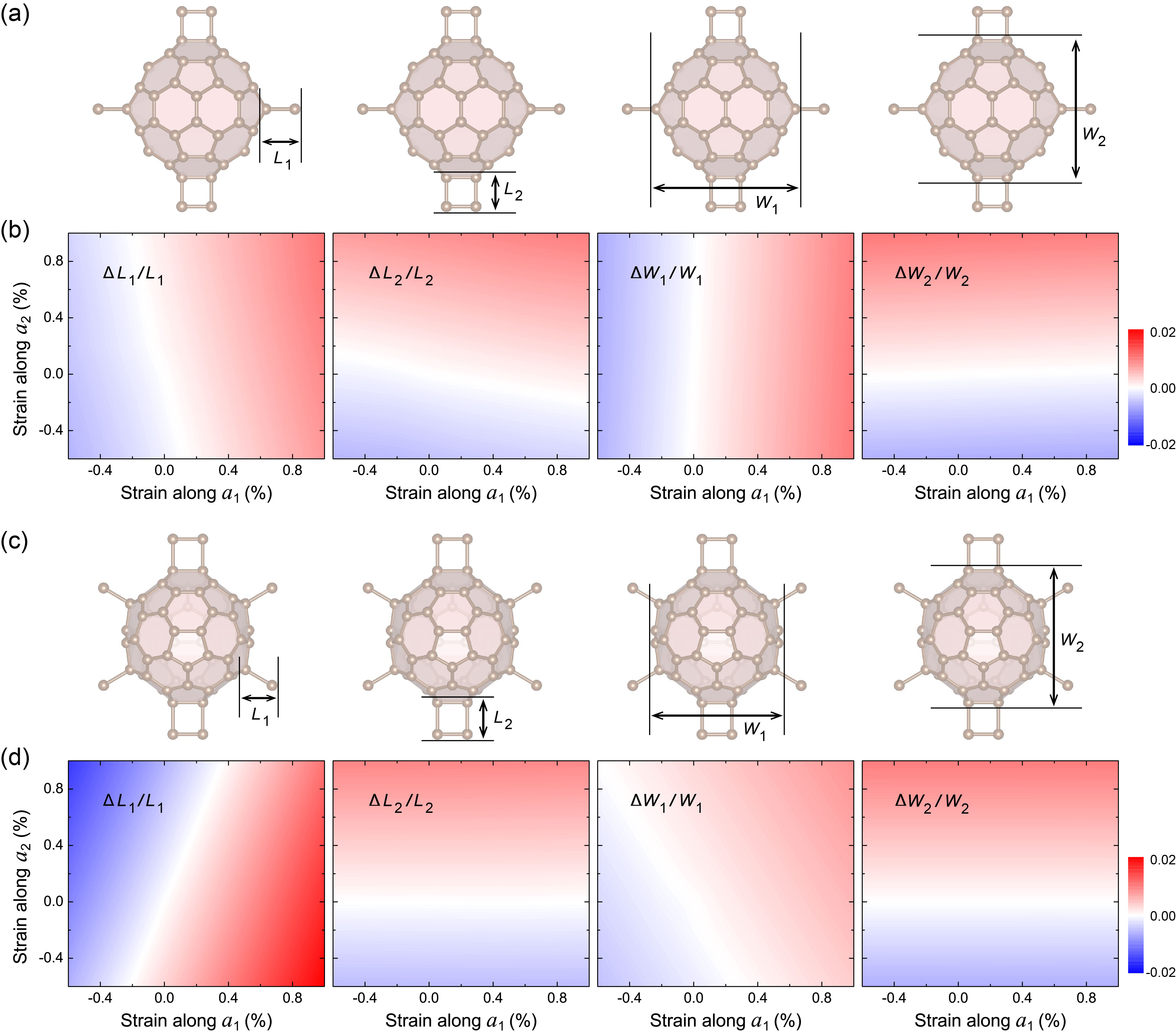}
\caption{
(a) Structural parameters of monolayer qTP fullerene networks and (b) their variations under strains. (c) Structural parameters of monolayer qHP fullerene networks and (d) their variations under strains.
}
\label{strain} 
\end{figure*}

Here we show that intermolecular bonds govern thermal expansion behaviors in fullerene networks. Interfullerene [2\,+\,2] cycloaddition bonds yield positive thermal expansion, whereas thermal contraction is found along single bonds. We identify the microscopic mechanism by analyzing the geometric flexibility of these bridge bonds. Remarkably, we find that the low-frequency transverse vibrations along the single bonds favor thermal contraction, in contrast to the transverse vibrations associated with the [2\,+\,2] cycloaddition bonds. Based on this understanding, we rationally design molecular networks beyond C$_{60}$ with tailored thermal expansion.

Thermal expansion is simulated under the quasi-harmonic approximation\,\cite{Huang2016,Pallikara2021} with volume-dependent phonons computed from density functional perturbation theory\,\cite{Gonze1995a,DFPT} using {\sc VASP}\,\cite{Kresse1996,Kresse1996a}. The Gibbs free energy is obtained by finding the unique minimum value of the Helmholtz free energy\,\cite{Dove1993,Togo2008,Togo2015} with varied lattice constants $a$ and $b$ at a strain step of 0.2\%\,\cite{Peng2019,Peng2023}.


Figure\,\ref{crystals}(a) shows the crystal structures of two distinct networks of C$_{60}$ monolayers, namely, the quasi-tetragonal phase (qTP) and quasi-hexagonal phase (qHP). For qTP, each carbon cage is connected by vertical and horizontal [2+2] cycloaddition bonds along the $a_1$ and $a_2$ directions, respectively. Such bonds are expected to expand rigidly along their axis upon heating. In qHP, only the buckyballs along $a_2$ are linked by the [2+2] cycloaddition bonds, while C$-$C single bonds link the neighboring cages along $a_1$. Figure\,\ref{crystals}(b) shows the thermal expansion of the two phases. While qTP along both directions has positive thermal expansion, negative thermal expansion along $a_1$ is found for qHP up to 500\,K, in contrast to the positive thermal expansion along $a_2$ (here we neglect the tiny negative thermal expansion along $a_2$ which might come from the soft vibrational modes of the isolated molecule with ellipsoidal deformation that keeps the surface area constant\,\cite{Kwon2004,Brown2006}). 

To understand the thermal expansion behaviors, we study the geometric flexibility of the two phases. Figure\,\ref{strain} shows the variations of structural parameters for qTP and qHP monolayer networks at varied strains along $a_1$ and $a_2$. These structural parameters measure the geometric flexibility of both intermolecular bonds ($L_{1,2}$) and individual molecules ($W_{1,2}$) along $a_1$ and $a_2$, as shown in Fig.\,\ref{strain}(a) and (c) for qTP and qHP, respectively. 

For qTP, both the intermolecular bond $L_1$ and the molecular width $W_1$ expand rigidly upon uniaxial strains along $a_1$. However, $\Delta L_1$ and $\Delta W_1$ remain nearly unchanged for strains along the other direction $a_2$, as shown by the color map in Fig.\,\ref{strain}(b). The same conclusion also holds for $L_2$ and $W_2$.

For qHP, the intermolecular [2\,+\,2] cycloaddition bonds $L_2$ and molecular width $W_2$ also expand rigidly for parallel strains along $a_2$, while resisting deformations for perpendicular strains along $a_1$, as shown in Fig.\,\ref{strain}(d), exhibiting behaviors similar to the [2\,+\,2] cycloaddition bonds in qTP. However, the intermolecular single bonds $L_1$ and molecular width $W_1$ in qHP respond differently to the strains. As shown by the $\Delta L_1$ color map in Fig.\,\ref{strain}(d), the $L_1$ in qHP C$_{60}$ becomes shorter with increased strain along $a_2$. This notable contraction along $L_1$ with increasing $a_2$ indicates that the single bonds deform more readily, with compression specifically along $a_1$ being favorable for positive strains along $a_2$. This hinge-like motion is expected in the single bonds as they are less resistant to perpendicular strains. On the other hand, the $W_1$ in qHP C$_{60}$ expands upon strains along $a_2$. Therefore, we can attribute negative thermal expansion in qHP to the geometric flexibility of the intermolecular single bonds instead of the molecules themselves.

The overall behavior demonstrates general features for different types of intermolecular bonds. The [2\,+\,2] cycloaddition bonds expand rigidly when strain is applied parallel to their direction while resisting deformations against perpendicular strains. Therefore, qTP C$_{60}$ networks exhibit nearly-isotropic positive thermal expansion. On the other hand, the single bonds allow for hinge-like compression when applying perpendicular strains. Unlike qTP, the flexible single bonds in qHP contract like hinges along $a_1$ when the rigid [2\,+\,2] cycloaddition bonds along $a_2$ expand upon heating. Our findings demonstrate a distinctive interplay between flexibility-driven lattice contraction via C$-$C single bonds and rigidity-induced structural expansion through [2\,+\,2] cycloaddition bonds. This drives a strong anisotropic thermal response in qHP C$_{60}$, as summarized by the schematics in Fig.\,\ref{crystals}(a). 

\begin{figure}
\centering
\includegraphics[width=0.88\linewidth]{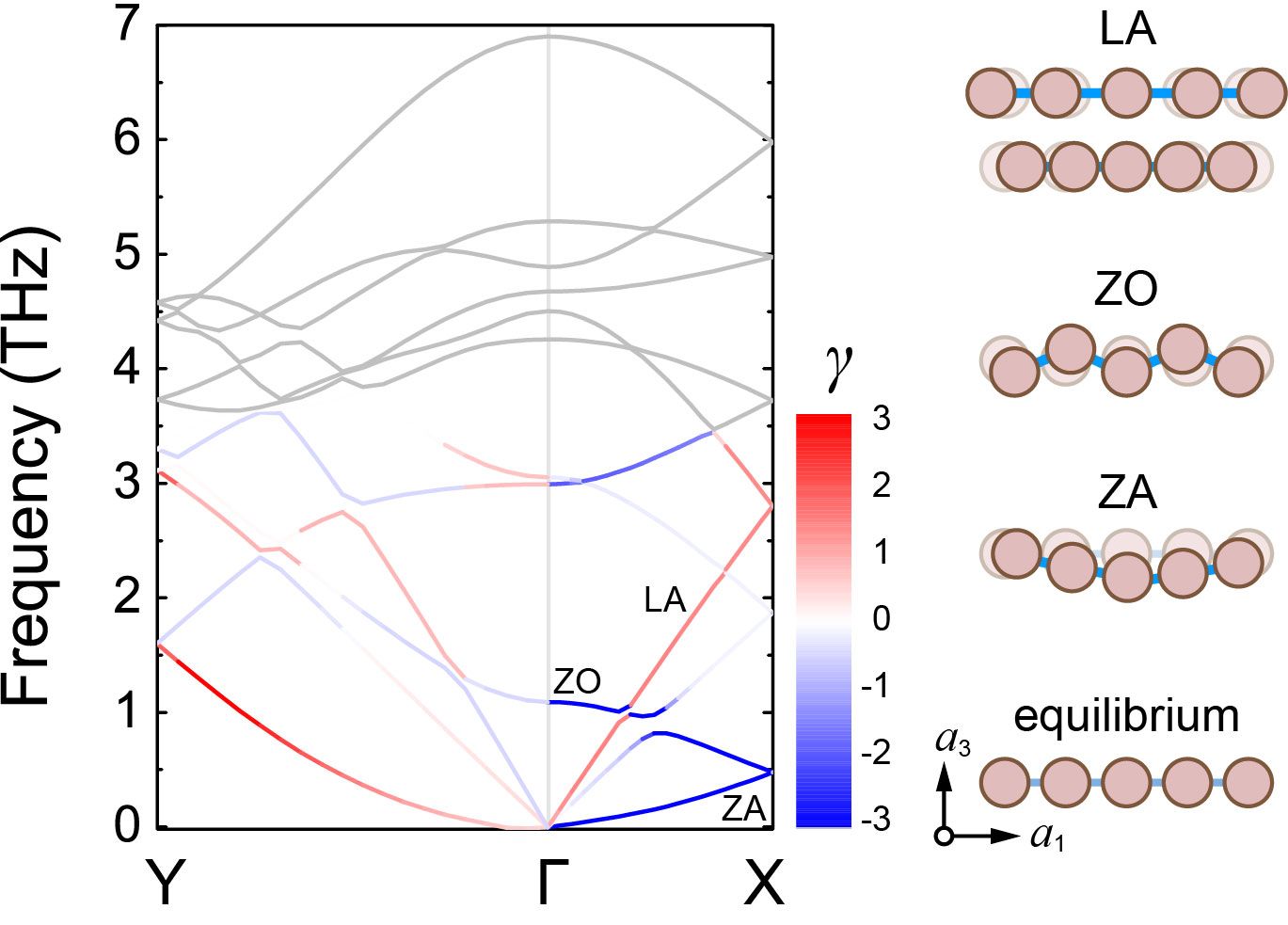}
\caption{
Low-frequency phonons, mode Gr\"uneisen parameters, and vibrational modes along $\Gamma$--X.
}
\label{phonon} 
\end{figure}

To further explore the microscopic mechanism of negative and positive anisotropic thermal expansion in qHP fullerene networks, we examine their vibrational modes. At 300 K, only phonons below 0.6 (4.3)\,THz have an occupation number above 10 (1) according to the Bose-Einstein distribution. Thus, we focus on low-frequency phonons hereafter. Figure\,\ref{phonon} shows the phonon dispersion curves and the corresponding mode Gr\"uneisen parameters $\gamma$. The phonon dispersion along Y--$\Gamma$ corresponds to the vibrations associated with the intermolecular [2\,+\,2] cycloaddition bonds along $a_2$. The mode Gr\"uneisen parameters of these vibrations are either positive or near zero, leading to thermal expansion. In contrast, the transverse displacements associated with the intermolecular single bonds have large negative $\gamma$ along $\Gamma$--X. There are two transverse phonon branches with the largest negative $\gamma$: an out-of-plane acoustic mode (ZA) with the coherent movement of all molecules along $z$, and an out-of-plane optical mode (ZO) with alternating displacements between neighboring molecules, as illustrated by the schematics in Fig.\,\ref{phonon}. Both modes with large negative $\gamma$ favor lattice contraction. In comparison, the longitudinal acoustic mode (LA) has positive $\gamma$, which contributes to lattice expansion instead but less strongly ($\gamma < 1.5$) than the ZA and ZO modes ($|\gamma| > 2.5$). The overall picture confirms that transverse displacements of the single bonds yield thermal contraction, while the [2\,+\,2] cycloaddition bond oscillations contribute to expansion behaviors.

The rigidity of the [2\,+\,2] cycloaddition bonds, as well as the flexibility of the C$-$C single bonds, provide a universal design principle to tailor thermal expansion behaviors. The [2\,+\,2] cycloaddition bonds between molecular cages impose structural, elastic, and vibrational constraints that only allow positive thermal expansion. Contrastingly, the flexible single bonds bridging the molecules allow for lattice contraction when perpendicular strains are applied, and the vibrational modes associated with these single bonds yield strong transverse displacements that favor thermal contraction. The flexibility of the single bonds is therefore the main driving factor in negative thermal expansion. Similar thermal behaviors in fullerene-based networks have also been found in previous molecular dynamics simulations\,\cite{Mortazavi2023}. Intuitively, we can either realize positive thermal expansion through intermolecular [2\,+\,2] cycloaddition bonds to resist lattice contraction, or utilize less rigid intermolecular single bonds to yield negative thermal expansion.

\begin{figure}
\centering
\includegraphics[width=0.88\linewidth]{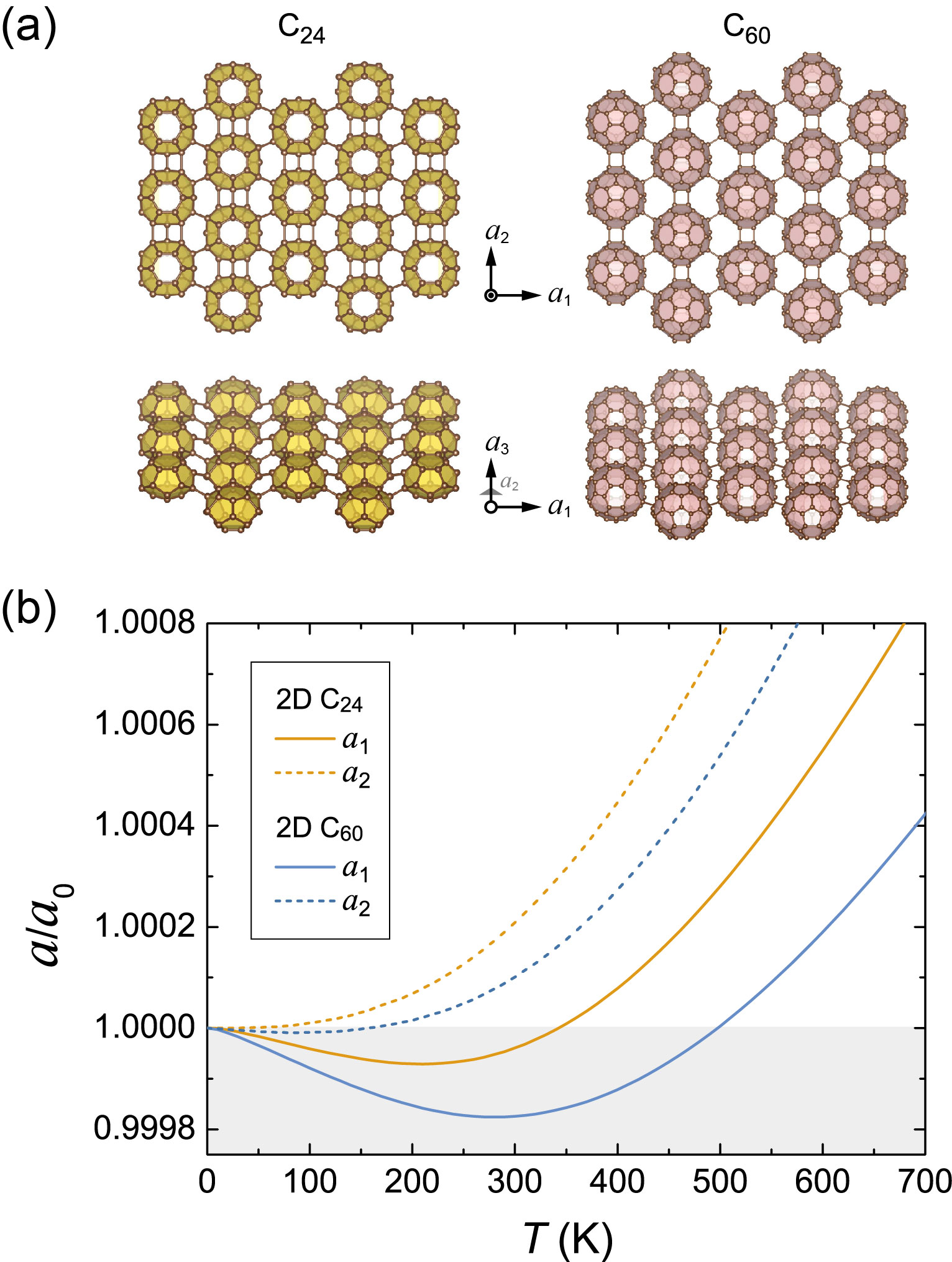}
\caption{
(a) Crystal structures and (b) thermal expansion of 2D qHP monolayer C$_{24}$ and C$_{60}$ networks. 
}
\label{C24} 
\end{figure}

The discovery that C$-$C single and [2\,+\,2] cycloaddition bonds contribute distinctively to thermal behavior provides a predictive tool for the rational design of thermally responsive materials with nearly-spherical building blocks such as icosahedral B$_{12}$ units\,\cite{Kah2015,Sun2017a,Zhang2017c}, fullerene cages\,\cite{Kroto1987,Hallett1995}, and all-metal clusters\,\cite{Xu2023} (for thermal expansion of 2D icosahedral B$_{12}$ networks, see Fig.\,\ref{boron} in the End Matter). As a proof-of-principle study, we extend this design principle to monolayer qHP C$_{24}$ networks, where the C$_{24}$ molecules are linked through similar single bonds along $a_1$ but three intermolecular bonds between six carbon atoms along $a_2$\,\cite{Wu2025}. The crystal structure of qHP C$_{24}$ is shown in Fig.\,\ref{C24}(a). Unlike the nearly-planar single bonds in qHP C$_{60}$, the single bonds in C$_{24}$ monolayers have a buckled structure owing to a larger molecular curvature and asymmetric intermolecular bonding positions. This, in combination with the smaller molecular size, leads to a higher density of interfullerene bonds and larger elastic constants than qHP C$_{60}$\,\cite{Wu2025}. It is therefore expected that the single bonds in C$_{24}$ are more rigid than those in qHP C$_{60}$, leading to stronger resistance to perpendicular deformations. Figure\,\ref{C24}(b) shows smaller negative thermal expansion in C$_{24}$ compared to C$_{60}$ along $a_1$ as expected. Additionally, the positive thermal expansion along $a_2$ in C$_{24}$ is also larger than that in C$_{60}$. This is unsurprising since the three intermolecular bonds along $a_2$ in C$_{24}$ are much more rigid than the [2\,+\,2] cycloaddition bonds in monolayer qHP C$_{60}$ networks.

The recent experimental realization of monolayer fullerene networks\,\cite{Hou2022} offers a timely opportunity to directly test our predictions, which influence numerous fullerene-based applications such as device integration in photodetectors\,\cite{Zhang2025} and catalytic activity in photocatalysis\,\cite{Peng2022c,Jones2023,Shearsby2025,Wang2023}. The temperature-dependent structural properties can be measured using high-resolution capacitance dilatometers\,\cite{Nagel1999} and synchrotron X-ray diffraction\,\cite{Arvanitidis2003}. The low-frequency phonon modes can be detected by Raman and infrared spectroscopy\,\cite{Brown2006}, enabling verification of the predicted transverse vibrational mechanisms. From an application perspective, molecular networks with tunable thermal expansion present compelling possibilities. Materials with designed negative or near-zero thermal expansion are in high demand in flexible electronics\,\cite{Mohn1999}, aerospace composites\,\cite{Strock1992}, and precision instrumentation\,\cite{Wang2024b}, where thermal mismatch must be minimized. Our results indicate that such properties can be achieved not through complex multi-phase composites, but through intrinsic molecular architecture. Previous experimental data on thermal expansion in molecular networks\,\cite{Baughman1993,Goodwin2008,Fortes2011,Dove2016,Liu2018b,Yao2019,Bond2021,Lee2021,Saha2024} including polymeric fullerene\,\cite{Blank1998,Nagel1999} confirm the fact that thermal expansion in molecular crystals can be controlled by bonding motifs and crystalline geometries, and similar mechanisms have been observed experimentally such as flexibility-driven ``hinge-like'' motion in crystalline networks\,\cite{Baughman1993,Goodwin2008,Fortes2011,Dove2016} and transverse displacements of bridging atoms\,\cite{Hancock2004,Goodwin2005} or membranes\,\cite{Huang2016c,Koocher2021}. This, again, supports the broader generalisability of our proposed mechanism.


In summary, we establish a general method for engineering thermal expansion in cage-like molecular networks through different intermolecular bonds. Using fullerene as a representative system, we show that intermolecular [2\,+\,2] cycloaddition bonds result in positive thermal expansion, whereas C$-$C single bonds lead to negative thermal expansion. By varying the structural frameworks of fullerene molecules, we can design a nearly-square lattice with positive thermal expansion through [2\,+\,2] cycloaddition bonds along both in-plane directions. Similarly, we can also realize negative and positive anisotropic thermal expansion in a nearly-triangular lattice through single bonds along one direction and [2\,+\,2] cycloaddition bonds along the other direction. The origin is related to both the geometric flexibility of different intermolecular bonds and their corresponding transverse vibrations. By identifying and uncovering this microscopic mechanism, we can further design molecular networks with tailored thermal expansion.


\begin{acknowledgements}
A.S. acknowledges support from Homerton College Cambridge for a Homerton-Victoria Brahm Schild Grant. J.W. acknowledges support from the Cambridge Undergraduate Research Opportunities Programme and from Peterhouse for the James Porter Scholarship. B.P. acknowledges support from Magdalene College Cambridge for a Nevile Research Fellowship. The calculations were performed using resources provided by the Cambridge Service for Data Driven Discovery (CSD3) operated by the University of Cambridge Research Computing Service (\url{www.csd3.cam.ac.uk}), provided by Dell EMC and Intel using Tier-2 funding from the Engineering and Physical Sciences Research Council (capital grant EP/T022159/1), and DiRAC funding from the Science and Technology Facilities Council (\url{http://www.dirac.ac.uk} ), as well as with computational support from the UK Materials and Molecular Modelling Hub, which is partially funded by EPSRC (EP/T022213/1, EP/W032260/1 and EP/P020194/1), for which access was obtained via the UKCP consortium and funded by EPSRC grant ref EP/P022561/1.
\end{acknowledgements}


\begin{thebibliography}{0}%
\makeatletter
\providecommand \@ifxundefined [1]{%
 \@ifx{#1\undefined}
}%
\providecommand \@ifnum [1]{%
 \ifnum #1\expandafter \@firstoftwo
 \else \expandafter \@secondoftwo
 \fi
}%
\providecommand \@ifx [1]{%
 \ifx #1\expandafter \@firstoftwo
 \else \expandafter \@secondoftwo
 \fi
}%
\providecommand \natexlab [1]{#1}%
\providecommand \enquote  [1]{``#1''}%
\providecommand \bibnamefont  [1]{#1}%
\providecommand \bibfnamefont [1]{#1}%
\providecommand \citenamefont [1]{#1}%
\providecommand \href@noop [0]{\@secondoftwo}%
\providecommand \href [0]{\begingroup \@sanitize@url \@href}%
\providecommand \@href[1]{\@@startlink{#1}\@@href}%
\providecommand \@@href[1]{\endgroup#1\@@endlink}%
\providecommand \@sanitize@url [0]{\catcode `\\12\catcode `\$12\catcode
  `\&12\catcode `\#12\catcode `\^12\catcode `\_12\catcode `\%12\relax}%
\providecommand \@@startlink[1]{}%
\providecommand \@@endlink[0]{}%
\providecommand \url  [0]{\begingroup\@sanitize@url \@url }%
\providecommand \@url [1]{\endgroup\@href {#1}{\urlprefix }}%
\providecommand \urlprefix  [0]{URL }%
\providecommand \Eprint [0]{\href }%
\providecommand \doibase [0]{http://dx.doi.org/}%
\providecommand \selectlanguage [0]{\@gobble}%
\providecommand \bibinfo  [0]{\@secondoftwo}%
\providecommand \bibfield  [0]{\@secondoftwo}%
\providecommand \translation [1]{[#1]}%
\providecommand \BibitemOpen [0]{}%
\providecommand \bibitemStop [0]{}%
\providecommand \bibitemNoStop [0]{.\EOS\space}%
\providecommand \EOS [0]{\spacefactor3000\relax}%
\providecommand \BibitemShut  [1]{\csname bibitem#1\endcsname}%
\let\auto@bib@innerbib\@empty
\end{thebibliography}%


\begin{thebibliography}{68}%
\makeatletter
\providecommand \@ifxundefined [1]{%
 \@ifx{#1\undefined}
}%
\providecommand \@ifnum [1]{%
 \ifnum #1\expandafter \@firstoftwo
 \else \expandafter \@secondoftwo
 \fi
}%
\providecommand \@ifx [1]{%
 \ifx #1\expandafter \@firstoftwo
 \else \expandafter \@secondoftwo
 \fi
}%
\providecommand \natexlab [1]{#1}%
\providecommand \enquote  [1]{``#1''}%
\providecommand \bibnamefont  [1]{#1}%
\providecommand \bibfnamefont [1]{#1}%
\providecommand \citenamefont [1]{#1}%
\providecommand \href@noop [0]{\@secondoftwo}%
\providecommand \href [0]{\begingroup \@sanitize@url \@href}%
\providecommand \@href[1]{\@@startlink{#1}\@@href}%
\providecommand \@@href[1]{\endgroup#1\@@endlink}%
\providecommand \@sanitize@url [0]{\catcode `\\12\catcode `\$12\catcode
  `\&12\catcode `\#12\catcode `\^12\catcode `\_12\catcode `\%12\relax}%
\providecommand \@@startlink[1]{}%
\providecommand \@@endlink[0]{}%
\providecommand \url  [0]{\begingroup\@sanitize@url \@url }%
\providecommand \@url [1]{\endgroup\@href {#1}{\urlprefix }}%
\providecommand \urlprefix  [0]{URL }%
\providecommand \Eprint [0]{\href }%
\providecommand \doibase [0]{http://dx.doi.org/}%
\providecommand \selectlanguage [0]{\@gobble}%
\providecommand \bibinfo  [0]{\@secondoftwo}%
\providecommand \bibfield  [0]{\@secondoftwo}%
\providecommand \translation [1]{[#1]}%
\providecommand \BibitemOpen [0]{}%
\providecommand \bibitemStop [0]{}%
\providecommand \bibitemNoStop [0]{.\EOS\space}%
\providecommand \EOS [0]{\spacefactor3000\relax}%
\providecommand \BibitemShut  [1]{\csname bibitem#1\endcsname}%
\let\auto@bib@innerbib\@empty
\bibitem [{\citenamefont {Johnson}\ and\ \citenamefont
  {Parsons}(1944)}]{Johnson1944}%
  \BibitemOpen
  \bibfield  {author} {\bibinfo {author} {\bibfnamefont {Walter~H}\
  \bibnamefont {Johnson}}\ and\ \bibinfo {author} {\bibfnamefont {Willard~H}\
  \bibnamefont {Parsons}},\ }\href@noop {} {\emph {\bibinfo {title} {Thermal
  expansion of concrete aggregate materials}}}\ (\bibinfo  {publisher} {US
  Government Printing Office Washington, DC, USA},\ \bibinfo {year}
  {1944})\BibitemShut {NoStop}%
\bibitem [{\citenamefont {Mohn}(1999)}]{Mohn1999}%
  \BibitemOpen
  \bibfield  {author} {\bibinfo {author} {\bibfnamefont {Peter}\ \bibnamefont
  {Mohn}},\ }\bibfield  {title} {\enquote {\bibinfo {title} {A century of zero
  expansion},}\ }\href {\doibase 10.1038/21778} {\bibfield  {journal} {\bibinfo
   {journal} {Nature}\ }\textbf {\bibinfo {volume} {400}},\ \bibinfo {pages}
  {18--19} (\bibinfo {year} {1999})}\BibitemShut {NoStop}%
\bibitem [{\citenamefont {Strock}(1992)}]{Strock1992}%
  \BibitemOpen
  \bibfield  {author} {\bibinfo {author} {\bibfnamefont {John~D.}\ \bibnamefont
  {Strock}},\ }\bibfield  {title} {\enquote {\bibinfo {title} {Development of
  zero coefficient of thermal expansion composite tubes for stable space
  structures},}\ }\href {\doibase 10.1117/12.137997} {\bibfield  {journal}
  {\bibinfo  {journal} {Proc. SPIE}\ }\textbf {\bibinfo {volume} {1690}},\
  \bibinfo {pages} {223--230} (\bibinfo {year} {1992})}\BibitemShut {NoStop}%
\bibitem [{\citenamefont {Toropova}\ and\ \citenamefont
  {Steeves}(2015)}]{Toropova2015}%
  \BibitemOpen
  \bibfield  {author} {\bibinfo {author} {\bibfnamefont {Marina~M.}\
  \bibnamefont {Toropova}}\ and\ \bibinfo {author} {\bibfnamefont {Craig~A.}\
  \bibnamefont {Steeves}},\ }\bibfield  {title} {\enquote {\bibinfo {title}
  {Adaptive bimaterial lattices to mitigate thermal expansion mismatch stresses
  in satellite structures},}\ }\href {\doibase 10.1016/j.actaastro.2015.03.022}
  {\bibfield  {journal} {\bibinfo  {journal} {Acta Astronautica}\ }\textbf
  {\bibinfo {volume} {113}},\ \bibinfo {pages} {132--141} (\bibinfo {year}
  {2015})}\BibitemShut {NoStop}%
\bibitem [{\citenamefont {Barrera}\ \emph {et~al.}(2005)\citenamefont
  {Barrera}, \citenamefont {Bruno}, \citenamefont {Barron},\ and\ \citenamefont
  {Allan}}]{Barrera2005}%
  \BibitemOpen
  \bibfield  {author} {\bibinfo {author} {\bibfnamefont {G~D}\ \bibnamefont
  {Barrera}}, \bibinfo {author} {\bibfnamefont {J~A~O}\ \bibnamefont {Bruno}},
  \bibinfo {author} {\bibfnamefont {T~H~K}\ \bibnamefont {Barron}}, \ and\
  \bibinfo {author} {\bibfnamefont {N~L}\ \bibnamefont {Allan}},\ }\bibfield
  {title} {\enquote {\bibinfo {title} {Negative thermal expansion},}\ }\href
  {\doibase 10.1088/0953-8984/17/4/R03} {\bibfield  {journal} {\bibinfo
  {journal} {Journal of Physics: Condensed Matter}\ }\textbf {\bibinfo {volume}
  {17}},\ \bibinfo {pages} {R217} (\bibinfo {year} {2005})}\BibitemShut
  {NoStop}%
\bibitem [{\citenamefont {Liu}\ \emph {et~al.}(2017)\citenamefont {Liu},
  \citenamefont {Shang},\ and\ \citenamefont {Wang}}]{Liu2017c}%
  \BibitemOpen
  \bibfield  {author} {\bibinfo {author} {\bibfnamefont {Zi-Kui}\ \bibnamefont
  {Liu}}, \bibinfo {author} {\bibfnamefont {Shun-Li}\ \bibnamefont {Shang}}, \
  and\ \bibinfo {author} {\bibfnamefont {Yi}~\bibnamefont {Wang}},\ }\bibfield
  {title} {\enquote {\bibinfo {title} {Fundamentals of thermal expansion and
  thermal contraction},}\ }\href {\doibase 10.3390/ma10040410} {\bibfield
  {journal} {\bibinfo  {journal} {Materials}\ }\textbf {\bibinfo {volume}
  {10}},\ \bibinfo {pages} {410} (\bibinfo {year} {2017})}\BibitemShut
  {NoStop}%
\bibitem [{\citenamefont {Baughman}\ and\ \citenamefont
  {Galv{\~a}o}(1993)}]{Baughman1993}%
  \BibitemOpen
  \bibfield  {author} {\bibinfo {author} {\bibfnamefont {Ray~H.}\ \bibnamefont
  {Baughman}}\ and\ \bibinfo {author} {\bibfnamefont {Douglas~S.}\ \bibnamefont
  {Galv{\~a}o}},\ }\bibfield  {title} {\enquote {\bibinfo {title} {Crystalline
  networks with unusual predicted mechanical and thermal properties},}\ }\href
  {\doibase 10.1038/365735a0} {\bibfield  {journal} {\bibinfo  {journal}
  {Nature}\ }\textbf {\bibinfo {volume} {365}},\ \bibinfo {pages} {735--737}
  (\bibinfo {year} {1993})}\BibitemShut {NoStop}%
\bibitem [{\citenamefont {Mary}\ \emph {et~al.}(1996)\citenamefont {Mary},
  \citenamefont {Evans}, \citenamefont {Vogt},\ and\ \citenamefont
  {Sleight}}]{Mary1996}%
  \BibitemOpen
  \bibfield  {author} {\bibinfo {author} {\bibfnamefont {T.~A.}\ \bibnamefont
  {Mary}}, \bibinfo {author} {\bibfnamefont {J.~S.~O.}\ \bibnamefont {Evans}},
  \bibinfo {author} {\bibfnamefont {T.}~\bibnamefont {Vogt}}, \ and\ \bibinfo
  {author} {\bibfnamefont {A.~W.}\ \bibnamefont {Sleight}},\ }\bibfield
  {title} {\enquote {\bibinfo {title} {Negative thermal expansion from 0.3 to
  1050 kelvin in zrw$_2$o$_8$},}\ }\href {\doibase 10.1126/science.272.5258.90}
  {\bibfield  {journal} {\bibinfo  {journal} {Science}\ }\textbf {\bibinfo
  {volume} {272}},\ \bibinfo {pages} {90--92} (\bibinfo {year}
  {1996})}\BibitemShut {NoStop}%
\bibitem [{\citenamefont {Das}\ \emph {et~al.}(2010)\citenamefont {Das},
  \citenamefont {Jacobs},\ and\ \citenamefont {Barbour}}]{Das2010}%
  \BibitemOpen
  \bibfield  {author} {\bibinfo {author} {\bibfnamefont {Dinabandhu}\
  \bibnamefont {Das}}, \bibinfo {author} {\bibfnamefont {Tia}\ \bibnamefont
  {Jacobs}}, \ and\ \bibinfo {author} {\bibfnamefont {Leonard~J.}\ \bibnamefont
  {Barbour}},\ }\bibfield  {title} {\enquote {\bibinfo {title} {Exceptionally
  large positive and negative anisotropic thermal expansion of an organic
  crystalline material},}\ }\href {\doibase 10.1038/nmat2583} {\bibfield
  {journal} {\bibinfo  {journal} {Nature Materials}\ }\textbf {\bibinfo
  {volume} {9}},\ \bibinfo {pages} {36--39} (\bibinfo {year}
  {2010})}\BibitemShut {NoStop}%
\bibitem [{\citenamefont {Fortes}\ \emph {et~al.}(2011)\citenamefont {Fortes},
  \citenamefont {Suard},\ and\ \citenamefont {Knight}}]{Fortes2011}%
  \BibitemOpen
  \bibfield  {author} {\bibinfo {author} {\bibfnamefont {A.~Dominic}\
  \bibnamefont {Fortes}}, \bibinfo {author} {\bibfnamefont {Emmanuelle}\
  \bibnamefont {Suard}}, \ and\ \bibinfo {author} {\bibfnamefont {Kevin~S.}\
  \bibnamefont {Knight}},\ }\bibfield  {title} {\enquote {\bibinfo {title}
  {Negative linear compressibility and massive anisotropic thermal expansion in
  methanol monohydrate},}\ }\href {\doibase 10.1126/science.1198640} {\bibfield
   {journal} {\bibinfo  {journal} {Science}\ }\textbf {\bibinfo {volume}
  {331}},\ \bibinfo {pages} {742--746} (\bibinfo {year} {2011})}\BibitemShut
  {NoStop}%
\bibitem [{\citenamefont {Takenaka}(2012)}]{Takenaka2012}%
  \BibitemOpen
  \bibfield  {author} {\bibinfo {author} {\bibfnamefont {Koshi}\ \bibnamefont
  {Takenaka}},\ }\bibfield  {title} {\enquote {\bibinfo {title} {Negative
  thermal expansion materials: technological key for control of thermal
  expansion},}\ }\href {\doibase 10.1088/1468-6996/13/1/013001} {\bibfield
  {journal} {\bibinfo  {journal} {Science and Technology of Advanced
  Materials}\ }\textbf {\bibinfo {volume} {13}},\ \bibinfo {pages} {013001}
  (\bibinfo {year} {2012})}\BibitemShut {NoStop}%
\bibitem [{\citenamefont {Liu}\ \emph {et~al.}(2018)\citenamefont {Liu},
  \citenamefont {Gao}, \citenamefont {Chen}, \citenamefont {Deng},
  \citenamefont {Lin},\ and\ \citenamefont {Xing}}]{Liu2018b}%
  \BibitemOpen
  \bibfield  {author} {\bibinfo {author} {\bibfnamefont {Zhanning}\
  \bibnamefont {Liu}}, \bibinfo {author} {\bibfnamefont {Qilong}\ \bibnamefont
  {Gao}}, \bibinfo {author} {\bibfnamefont {Jun}\ \bibnamefont {Chen}},
  \bibinfo {author} {\bibfnamefont {Jinxia}\ \bibnamefont {Deng}}, \bibinfo
  {author} {\bibfnamefont {Kun}\ \bibnamefont {Lin}}, \ and\ \bibinfo {author}
  {\bibfnamefont {Xianran}\ \bibnamefont {Xing}},\ }\bibfield  {title}
  {\enquote {\bibinfo {title} {Negative thermal expansion in molecular
  materials},}\ }\href {\doibase 10.1039/C8CC01153B} {\bibfield  {journal}
  {\bibinfo  {journal} {Chem. Commun.}\ }\textbf {\bibinfo {volume} {54}},\
  \bibinfo {pages} {5164--5176} (\bibinfo {year} {2018})}\BibitemShut {NoStop}%
\bibitem [{\citenamefont {Ritz}\ and\ \citenamefont
  {Benedek}(2018)}]{Ritz2018}%
  \BibitemOpen
  \bibfield  {author} {\bibinfo {author} {\bibfnamefont {Ethan~T.}\
  \bibnamefont {Ritz}}\ and\ \bibinfo {author} {\bibfnamefont {Nicole~A.}\
  \bibnamefont {Benedek}},\ }\bibfield  {title} {\enquote {\bibinfo {title}
  {Interplay between phonons and anisotropic elasticity drives negative thermal
  expansion in pbtio$_{3}$},}\ }\href {\doibase 10.1103/PhysRevLett.121.255901}
  {\bibfield  {journal} {\bibinfo  {journal} {Phys. Rev. Lett.}\ }\textbf
  {\bibinfo {volume} {121}},\ \bibinfo {pages} {255901} (\bibinfo {year}
  {2018})}\BibitemShut {NoStop}%
\bibitem [{\citenamefont {Shi}\ \emph {et~al.}(2021)\citenamefont {Shi},
  \citenamefont {Song}, \citenamefont {Xing},\ and\ \citenamefont
  {Chen}}]{Shi2021}%
  \BibitemOpen
  \bibfield  {author} {\bibinfo {author} {\bibfnamefont {Naike}\ \bibnamefont
  {Shi}}, \bibinfo {author} {\bibfnamefont {Yuzhu}\ \bibnamefont {Song}},
  \bibinfo {author} {\bibfnamefont {Xianran}\ \bibnamefont {Xing}}, \ and\
  \bibinfo {author} {\bibfnamefont {Jun}\ \bibnamefont {Chen}},\ }\bibfield
  {title} {\enquote {\bibinfo {title} {Negative thermal expansion in framework
  structure materials},}\ }\href {\doibase 10.1016/j.ccr.2021.214204}
  {\bibfield  {journal} {\bibinfo  {journal} {Coordination Chemistry Reviews}\
  }\textbf {\bibinfo {volume} {449}},\ \bibinfo {pages} {214204--} (\bibinfo
  {year} {2021})}\BibitemShut {NoStop}%
\bibitem [{\citenamefont {Li}\ \emph {et~al.}(2022)\citenamefont {Li},
  \citenamefont {Lin}, \citenamefont {Liu}, \citenamefont {Hu}, \citenamefont
  {Cao}, \citenamefont {Chen},\ and\ \citenamefont {Xing}}]{Li2022a}%
  \BibitemOpen
  \bibfield  {author} {\bibinfo {author} {\bibfnamefont {Qiang}\ \bibnamefont
  {Li}}, \bibinfo {author} {\bibfnamefont {Kun}\ \bibnamefont {Lin}}, \bibinfo
  {author} {\bibfnamefont {Zhanning}\ \bibnamefont {Liu}}, \bibinfo {author}
  {\bibfnamefont {Lei}\ \bibnamefont {Hu}}, \bibinfo {author} {\bibfnamefont
  {Yili}\ \bibnamefont {Cao}}, \bibinfo {author} {\bibfnamefont {Jun}\
  \bibnamefont {Chen}}, \ and\ \bibinfo {author} {\bibfnamefont {Xianran}\
  \bibnamefont {Xing}},\ }\bibfield  {title} {\enquote {\bibinfo {title}
  {Chemical diversity for tailoring negative thermal expansion},}\ }\href
  {\doibase 10.1021/acs.chemrev.1c00756} {\bibfield  {journal} {\bibinfo
  {journal} {Chem. Rev.}\ }\textbf {\bibinfo {volume} {122}},\ \bibinfo {pages}
  {8438--8486} (\bibinfo {year} {2022})}\BibitemShut {NoStop}%
\bibitem [{\citenamefont {Goodwin}\ \emph {et~al.}(2008)\citenamefont
  {Goodwin}, \citenamefont {Calleja}, \citenamefont {Conterio}, \citenamefont
  {Dove}, \citenamefont {Evans}, \citenamefont {Keen}, \citenamefont {Peters},\
  and\ \citenamefont {Tucker}}]{Goodwin2008}%
  \BibitemOpen
  \bibfield  {author} {\bibinfo {author} {\bibfnamefont {Andrew~L.}\
  \bibnamefont {Goodwin}}, \bibinfo {author} {\bibfnamefont {Mark}\
  \bibnamefont {Calleja}}, \bibinfo {author} {\bibfnamefont {Michael~J.}\
  \bibnamefont {Conterio}}, \bibinfo {author} {\bibfnamefont {Martin~T.}\
  \bibnamefont {Dove}}, \bibinfo {author} {\bibfnamefont {John S.~O.}\
  \bibnamefont {Evans}}, \bibinfo {author} {\bibfnamefont {David~A.}\
  \bibnamefont {Keen}}, \bibinfo {author} {\bibfnamefont {Lars}\ \bibnamefont
  {Peters}}, \ and\ \bibinfo {author} {\bibfnamefont {Matthew~G.}\ \bibnamefont
  {Tucker}},\ }\bibfield  {title} {\enquote {\bibinfo {title} {Colossal
  positive and negative thermal expansion in the framework material
  ag$_3$[co(cn)$_6$]},}\ }\href {\doibase 10.1126/science.1151442} {\bibfield
  {journal} {\bibinfo  {journal} {Science}\ }\textbf {\bibinfo {volume}
  {319}},\ \bibinfo {pages} {794--797} (\bibinfo {year} {2008})}\BibitemShut
  {NoStop}%
\bibitem [{\citenamefont {Dove}\ and\ \citenamefont {Fang}(2016)}]{Dove2016}%
  \BibitemOpen
  \bibfield  {author} {\bibinfo {author} {\bibfnamefont {Martin~T}\
  \bibnamefont {Dove}}\ and\ \bibinfo {author} {\bibfnamefont {Hong}\
  \bibnamefont {Fang}},\ }\bibfield  {title} {\enquote {\bibinfo {title}
  {Negative thermal expansion and associated anomalous physical properties:
  review of the lattice dynamics theoretical foundation},}\ }\href {\doibase
  10.1088/0034-4885/79/6/066503} {\bibfield  {journal} {\bibinfo  {journal}
  {Reports on Progress in Physics}\ }\textbf {\bibinfo {volume} {79}},\
  \bibinfo {pages} {066503} (\bibinfo {year} {2016})}\BibitemShut {NoStop}%
\bibitem [{\citenamefont {Pryde}\ \emph {et~al.}(1996)\citenamefont {Pryde},
  \citenamefont {Hammonds}, \citenamefont {Dove}, \citenamefont {Heine},
  \citenamefont {Gale},\ and\ \citenamefont {Warren}}]{Pryde1996}%
  \BibitemOpen
  \bibfield  {author} {\bibinfo {author} {\bibfnamefont {Alexandra K~A}\
  \bibnamefont {Pryde}}, \bibinfo {author} {\bibfnamefont {Kenton~D}\
  \bibnamefont {Hammonds}}, \bibinfo {author} {\bibfnamefont {Martin~T}\
  \bibnamefont {Dove}}, \bibinfo {author} {\bibfnamefont {Volker}\ \bibnamefont
  {Heine}}, \bibinfo {author} {\bibfnamefont {Julian~D}\ \bibnamefont {Gale}},
  \ and\ \bibinfo {author} {\bibfnamefont {Michele~C}\ \bibnamefont {Warren}},\
  }\bibfield  {title} {\enquote {\bibinfo {title} {{{Origin of the negative
  thermal expansion in ZrW$_2$O$_8$ and ZrV$_2$O$_7$}}},}\ }\href {\doibase
  10.1088/0953-8984/8/50/023} {\bibfield  {journal} {\bibinfo  {journal}
  {Journal of Physics: Condensed Matter}\ }\textbf {\bibinfo {volume} {8}},\
  \bibinfo {pages} {10973} (\bibinfo {year} {1996})}\BibitemShut {NoStop}%
\bibitem [{\citenamefont {Sleight}(1998)}]{Sleight1998}%
  \BibitemOpen
  \bibfield  {author} {\bibinfo {author} {\bibfnamefont {A.~W.}\ \bibnamefont
  {Sleight}},\ }\bibfield  {title} {\enquote {\bibinfo {title} {Compounds that
  contract on heating},}\ }\href {\doibase 10.1021/ic980253h} {\bibfield
  {journal} {\bibinfo  {journal} {Inorg. Chem.}\ }\textbf {\bibinfo {volume}
  {37}},\ \bibinfo {pages} {2854--2860} (\bibinfo {year} {1998})}\BibitemShut
  {NoStop}%
\bibitem [{\citenamefont {Tucker}\ \emph {et~al.}(2005)\citenamefont {Tucker},
  \citenamefont {Goodwin}, \citenamefont {Dove}, \citenamefont {Keen},
  \citenamefont {Wells},\ and\ \citenamefont {Evans}}]{Tucker2005}%
  \BibitemOpen
  \bibfield  {author} {\bibinfo {author} {\bibfnamefont {Matthew~G.}\
  \bibnamefont {Tucker}}, \bibinfo {author} {\bibfnamefont {Andrew~L.}\
  \bibnamefont {Goodwin}}, \bibinfo {author} {\bibfnamefont {Martin~T.}\
  \bibnamefont {Dove}}, \bibinfo {author} {\bibfnamefont {David~A.}\
  \bibnamefont {Keen}}, \bibinfo {author} {\bibfnamefont {Stephen~A.}\
  \bibnamefont {Wells}}, \ and\ \bibinfo {author} {\bibfnamefont {John S.~O.}\
  \bibnamefont {Evans}},\ }\bibfield  {title} {\enquote {\bibinfo {title}
  {Negative thermal expansion in zrw$_{2}$o$_{8}$: Mechanisms, rigid unit
  modes, and neutron total scattering},}\ }\href {\doibase
  10.1103/PhysRevLett.95.255501} {\bibfield  {journal} {\bibinfo  {journal}
  {Phys. Rev. Lett.}\ }\textbf {\bibinfo {volume} {95}},\ \bibinfo {pages}
  {255501} (\bibinfo {year} {2005})}\BibitemShut {NoStop}%
\bibitem [{\citenamefont {Tan}\ \emph {et~al.}(2024)\citenamefont {Tan},
  \citenamefont {Heine}, \citenamefont {Li},\ and\ \citenamefont
  {Dove}}]{Tan2024}%
  \BibitemOpen
  \bibfield  {author} {\bibinfo {author} {\bibfnamefont {Lei}\ \bibnamefont
  {Tan}}, \bibinfo {author} {\bibfnamefont {Volker}\ \bibnamefont {Heine}},
  \bibinfo {author} {\bibfnamefont {Gong}\ \bibnamefont {Li}}, \ and\ \bibinfo
  {author} {\bibfnamefont {Martin~T}\ \bibnamefont {Dove}},\ }\bibfield
  {title} {\enquote {\bibinfo {title} {The rigid unit mode model: review of
  ideas and applications},}\ }\href {\doibase 10.1088/1361-6633/acc7b7}
  {\bibfield  {journal} {\bibinfo  {journal} {Reports on Progress in Physics}\
  }\textbf {\bibinfo {volume} {87}},\ \bibinfo {pages} {126501} (\bibinfo
  {year} {2024})}\BibitemShut {NoStop}%
\bibitem [{\citenamefont {Hancock}\ \emph {et~al.}(2004)\citenamefont
  {Hancock}, \citenamefont {Turpen}, \citenamefont {Schlesinger}, \citenamefont
  {Kowach},\ and\ \citenamefont {Ramirez}}]{Hancock2004}%
  \BibitemOpen
  \bibfield  {author} {\bibinfo {author} {\bibfnamefont {Jason~N.}\
  \bibnamefont {Hancock}}, \bibinfo {author} {\bibfnamefont {Chandra}\
  \bibnamefont {Turpen}}, \bibinfo {author} {\bibfnamefont {Zack}\ \bibnamefont
  {Schlesinger}}, \bibinfo {author} {\bibfnamefont {Glen~R.}\ \bibnamefont
  {Kowach}}, \ and\ \bibinfo {author} {\bibfnamefont {Arthur~P.}\ \bibnamefont
  {Ramirez}},\ }\bibfield  {title} {\enquote {\bibinfo {title} {Unusual
  low-energy phonon dynamics in the negative thermal expansion compound
  zrw$_{2}$o$_{8}$},}\ }\href {\doibase 10.1103/PhysRevLett.93.225501}
  {\bibfield  {journal} {\bibinfo  {journal} {Phys. Rev. Lett.}\ }\textbf
  {\bibinfo {volume} {93}},\ \bibinfo {pages} {225501} (\bibinfo {year}
  {2004})}\BibitemShut {NoStop}%
\bibitem [{\citenamefont {Goodwin}\ and\ \citenamefont
  {Kepert}(2005)}]{Goodwin2005}%
  \BibitemOpen
  \bibfield  {author} {\bibinfo {author} {\bibfnamefont {Andrew~L.}\
  \bibnamefont {Goodwin}}\ and\ \bibinfo {author} {\bibfnamefont {Cameron~J.}\
  \bibnamefont {Kepert}},\ }\bibfield  {title} {\enquote {\bibinfo {title}
  {Negative thermal expansion and low-frequency modes in cyanide-bridged
  framework materials},}\ }\href {\doibase 10.1103/PhysRevB.71.140301}
  {\bibfield  {journal} {\bibinfo  {journal} {Phys. Rev. B}\ }\textbf {\bibinfo
  {volume} {71}},\ \bibinfo {pages} {140301} (\bibinfo {year}
  {2005})}\BibitemShut {NoStop}%
\bibitem [{\citenamefont {Huang}\ \emph
  {et~al.}(2016{\natexlab{a}})\citenamefont {Huang}, \citenamefont {Lu},\ and\
  \citenamefont {Rondinelli}}]{Huang2016c}%
  \BibitemOpen
  \bibfield  {author} {\bibinfo {author} {\bibfnamefont {Liang-Feng}\
  \bibnamefont {Huang}}, \bibinfo {author} {\bibfnamefont {Xue-Zeng}\
  \bibnamefont {Lu}}, \ and\ \bibinfo {author} {\bibfnamefont {James~M.}\
  \bibnamefont {Rondinelli}},\ }\bibfield  {title} {\enquote {\bibinfo {title}
  {Tunable negative thermal expansion in layered perovskites from
  quasi-two-dimensional vibrations},}\ }\href {\doibase
  10.1103/PhysRevLett.117.115901} {\bibfield  {journal} {\bibinfo  {journal}
  {Phys. Rev. Lett.}\ }\textbf {\bibinfo {volume} {117}},\ \bibinfo {pages}
  {115901} (\bibinfo {year} {2016}{\natexlab{a}})}\BibitemShut {NoStop}%
\bibitem [{\citenamefont {Koocher}\ \emph {et~al.}(2021)\citenamefont
  {Koocher}, \citenamefont {Huang},\ and\ \citenamefont
  {Rondinelli}}]{Koocher2021}%
  \BibitemOpen
  \bibfield  {author} {\bibinfo {author} {\bibfnamefont {Nathan~Z.}\
  \bibnamefont {Koocher}}, \bibinfo {author} {\bibfnamefont {Liang-Feng}\
  \bibnamefont {Huang}}, \ and\ \bibinfo {author} {\bibfnamefont {James~M.}\
  \bibnamefont {Rondinelli}},\ }\bibfield  {title} {\enquote {\bibinfo {title}
  {Negative thermal expansion in the ruddlesden-popper calcium titanates},}\
  }\href {\doibase 10.1103/PhysRevMaterials.5.053601} {\bibfield  {journal}
  {\bibinfo  {journal} {Phys. Rev. Mater.}\ }\textbf {\bibinfo {volume} {5}},\
  \bibinfo {pages} {053601} (\bibinfo {year} {2021})}\BibitemShut {NoStop}%
\bibitem [{\citenamefont {Hou}\ \emph {et~al.}(2022)\citenamefont {Hou},
  \citenamefont {Cui}, \citenamefont {Guan}, \citenamefont {Wang},
  \citenamefont {Li}, \citenamefont {Liu}, \citenamefont {Zhu},\ and\
  \citenamefont {Zheng}}]{Hou2022}%
  \BibitemOpen
  \bibfield  {author} {\bibinfo {author} {\bibfnamefont {Lingxiang}\
  \bibnamefont {Hou}}, \bibinfo {author} {\bibfnamefont {Xueping}\ \bibnamefont
  {Cui}}, \bibinfo {author} {\bibfnamefont {Bo}~\bibnamefont {Guan}}, \bibinfo
  {author} {\bibfnamefont {Shaozhi}\ \bibnamefont {Wang}}, \bibinfo {author}
  {\bibfnamefont {Ruian}\ \bibnamefont {Li}}, \bibinfo {author} {\bibfnamefont
  {Yunqi}\ \bibnamefont {Liu}}, \bibinfo {author} {\bibfnamefont {Daoben}\
  \bibnamefont {Zhu}}, \ and\ \bibinfo {author} {\bibfnamefont {Jian}\
  \bibnamefont {Zheng}},\ }\bibfield  {title} {\enquote {\bibinfo {title}
  {Synthesis of a monolayer fullerene network},}\ }\href {\doibase
  10.1038/s41586-022-04771-5} {\bibfield  {journal} {\bibinfo  {journal}
  {Nature}\ }\textbf {\bibinfo {volume} {606}},\ \bibinfo {pages} {507--510}
  (\bibinfo {year} {2022})}\BibitemShut {NoStop}%
\bibitem [{\citenamefont {Peng}(2022)}]{Peng2022c}%
  \BibitemOpen
  \bibfield  {author} {\bibinfo {author} {\bibfnamefont {Bo}~\bibnamefont
  {Peng}},\ }\bibfield  {title} {\enquote {\bibinfo {title} {Monolayer
  fullerene networks as photocatalysts for overall water splitting},}\ }\href
  {\doibase 10.1021/jacs.2c08054} {\bibfield  {journal} {\bibinfo  {journal}
  {J. Am. Chem. Soc.}\ }\textbf {\bibinfo {volume} {144}},\ \bibinfo {pages}
  {19921--19931} (\bibinfo {year} {2022})}\BibitemShut {NoStop}%
\bibitem [{\citenamefont {Peng}(2023)}]{Peng2023}%
  \BibitemOpen
  \bibfield  {author} {\bibinfo {author} {\bibfnamefont {Bo}~\bibnamefont
  {Peng}},\ }\bibfield  {title} {\enquote {\bibinfo {title} {Stability and
  strength of monolayer polymeric c$_{60}$},}\ }\href {\doibase
  10.1021/acs.nanolett.2c04497} {\bibfield  {journal} {\bibinfo  {journal}
  {Nano Lett.}\ }\textbf {\bibinfo {volume} {23}},\ \bibinfo {pages} {652--658}
  (\bibinfo {year} {2023})}\BibitemShut {NoStop}%
\bibitem [{\citenamefont {Jones}\ and\ \citenamefont {Peng}(2023)}]{Jones2023}%
  \BibitemOpen
  \bibfield  {author} {\bibinfo {author} {\bibfnamefont {Cory}\ \bibnamefont
  {Jones}}\ and\ \bibinfo {author} {\bibfnamefont {Bo}~\bibnamefont {Peng}},\
  }\bibfield  {title} {\enquote {\bibinfo {title} {Boosting photocatalytic
  water splitting of polymeric c60 by reduced dimensionality from
  two-dimensional monolayer to one-dimensional chain},}\ }\href {\doibase
  10.1021/acs.jpclett.3c02578} {\bibfield  {journal} {\bibinfo  {journal} {J.
  Phys. Chem. Lett.}\ }\textbf {\bibinfo {volume} {14}},\ \bibinfo {pages}
  {11768--11773} (\bibinfo {year} {2023})}\BibitemShut {NoStop}%
\bibitem [{\citenamefont {Shearsby}\ \emph {et~al.}(2025)\citenamefont
  {Shearsby}, \citenamefont {Wu}, \citenamefont {Yang},\ and\ \citenamefont
  {Peng}}]{Shearsby2025}%
  \BibitemOpen
  \bibfield  {author} {\bibinfo {author} {\bibfnamefont {Dylan}\ \bibnamefont
  {Shearsby}}, \bibinfo {author} {\bibfnamefont {Jiaqi}\ \bibnamefont {Wu}},
  \bibinfo {author} {\bibfnamefont {Dekun}\ \bibnamefont {Yang}}, \ and\
  \bibinfo {author} {\bibfnamefont {Bo}~\bibnamefont {Peng}},\ }\bibfield
  {title} {\enquote {\bibinfo {title} {Tuning electronic and optical properties
  of 2d polymeric c60 by stacking two layers},}\ }\href {\doibase
  10.1039/D4NR04540H} {\bibfield  {journal} {\bibinfo  {journal} {Nanoscale}\
  }\textbf {\bibinfo {volume} {17}},\ \bibinfo {pages} {2616--2620} (\bibinfo
  {year} {2025})}\BibitemShut {NoStop}%
\bibitem [{\citenamefont {Kayley}\ and\ \citenamefont
  {Peng}(2025)}]{Kayley2025}%
  \BibitemOpen
  \bibfield  {author} {\bibinfo {author} {\bibfnamefont {Darius}\ \bibnamefont
  {Kayley}}\ and\ \bibinfo {author} {\bibfnamefont {Bo}~\bibnamefont {Peng}},\
  }\bibfield  {title} {\enquote {\bibinfo {title} {C$_{60}$ building blocks
  with tuneable structures for tailored functionalities},}\ }\href {\doibase
  10.1016/j.commt.2025.100030} {\bibfield  {journal} {\bibinfo  {journal}
  {Computational Materials Today}\ }\textbf {\bibinfo {volume} {6}},\ \bibinfo
  {pages} {100030} (\bibinfo {year} {2025})}\BibitemShut {NoStop}%
\bibitem [{\citenamefont {Peng}\ and\ \citenamefont
  {Pizzochero}(2025{\natexlab{a}})}]{Peng2025a}%
  \BibitemOpen
  \bibfield  {author} {\bibinfo {author} {\bibfnamefont {Bo}~\bibnamefont
  {Peng}}\ and\ \bibinfo {author} {\bibfnamefont {Michele}\ \bibnamefont
  {Pizzochero}},\ }\bibfield  {title} {\enquote {\bibinfo {title} {{{Monolayer
  C$_{60}$ networks: a first-principles perspective}}},}\ }\href {\doibase
  10.1039/D5CC02473K} {\bibfield  {journal} {\bibinfo  {journal} {Chem.
  Commun.}\ }\textbf {\bibinfo {volume} {61}},\ \bibinfo {pages} {10287--10302}
  (\bibinfo {year} {2025}{\natexlab{a}})}\BibitemShut {NoStop}%
\bibitem [{\citenamefont {Peng}\ and\ \citenamefont
  {Pizzochero}(2025{\natexlab{b}})}]{Peng2025c}%
  \BibitemOpen
  \bibfield  {author} {\bibinfo {author} {\bibfnamefont {Bo}~\bibnamefont
  {Peng}}\ and\ \bibinfo {author} {\bibfnamefont {Michele}\ \bibnamefont
  {Pizzochero}},\ }\bibfield  {title} {\enquote {\bibinfo {title} {Electronic
  structure of fullerene nanoribbons},}\ }\href {\doibase
  10.1021/acsnano.5c08991} {\bibfield  {journal} {\bibinfo  {journal} {ACS
  Nano}\ } (\bibinfo {year} {2025}{\natexlab{b}}),\
  10.1021/acsnano.5c08991}\BibitemShut {NoStop}%
\bibitem [{\citenamefont {Meirzadeh}\ \emph {et~al.}(2023)\citenamefont
  {Meirzadeh}, \citenamefont {Evans}, \citenamefont {Rezaee}, \citenamefont
  {Milich}, \citenamefont {Dionne}, \citenamefont {Darlington}, \citenamefont
  {Bao}, \citenamefont {Bartholomew}, \citenamefont {Handa}, \citenamefont
  {Rizzo}, \citenamefont {Wiscons}, \citenamefont {Reza}, \citenamefont
  {Zangiabadi}, \citenamefont {Fardian-Melamed}, \citenamefont {Crowther},
  \citenamefont {Schuck}, \citenamefont {Basov}, \citenamefont {Zhu},
  \citenamefont {Giri}, \citenamefont {Hopkins}, \citenamefont {Kim},
  \citenamefont {Steigerwald}, \citenamefont {Yang}, \citenamefont {Nuckolls},\
  and\ \citenamefont {Roy}}]{Meirzadeh2023}%
  \BibitemOpen
  \bibfield  {author} {\bibinfo {author} {\bibfnamefont {Elena}\ \bibnamefont
  {Meirzadeh}}, \bibinfo {author} {\bibfnamefont {Austin~M.}\ \bibnamefont
  {Evans}}, \bibinfo {author} {\bibfnamefont {Mehdi}\ \bibnamefont {Rezaee}},
  \bibinfo {author} {\bibfnamefont {Milena}\ \bibnamefont {Milich}}, \bibinfo
  {author} {\bibfnamefont {Connor~J.}\ \bibnamefont {Dionne}}, \bibinfo
  {author} {\bibfnamefont {Thomas~P.}\ \bibnamefont {Darlington}}, \bibinfo
  {author} {\bibfnamefont {Si~Tong}\ \bibnamefont {Bao}}, \bibinfo {author}
  {\bibfnamefont {Amymarie~K.}\ \bibnamefont {Bartholomew}}, \bibinfo {author}
  {\bibfnamefont {Taketo}\ \bibnamefont {Handa}}, \bibinfo {author}
  {\bibfnamefont {Daniel~J.}\ \bibnamefont {Rizzo}}, \bibinfo {author}
  {\bibfnamefont {Ren~A.}\ \bibnamefont {Wiscons}}, \bibinfo {author}
  {\bibfnamefont {Mahniz}\ \bibnamefont {Reza}}, \bibinfo {author}
  {\bibfnamefont {Amirali}\ \bibnamefont {Zangiabadi}}, \bibinfo {author}
  {\bibfnamefont {Natalie}\ \bibnamefont {Fardian-Melamed}}, \bibinfo {author}
  {\bibfnamefont {Andrew~C.}\ \bibnamefont {Crowther}}, \bibinfo {author}
  {\bibfnamefont {P.~James}\ \bibnamefont {Schuck}}, \bibinfo {author}
  {\bibfnamefont {D.~N.}\ \bibnamefont {Basov}}, \bibinfo {author}
  {\bibfnamefont {Xiaoyang}\ \bibnamefont {Zhu}}, \bibinfo {author}
  {\bibfnamefont {Ashutosh}\ \bibnamefont {Giri}}, \bibinfo {author}
  {\bibfnamefont {Patrick~E.}\ \bibnamefont {Hopkins}}, \bibinfo {author}
  {\bibfnamefont {Philip}\ \bibnamefont {Kim}}, \bibinfo {author}
  {\bibfnamefont {Michael~L.}\ \bibnamefont {Steigerwald}}, \bibinfo {author}
  {\bibfnamefont {Jingjing}\ \bibnamefont {Yang}}, \bibinfo {author}
  {\bibfnamefont {Colin}\ \bibnamefont {Nuckolls}}, \ and\ \bibinfo {author}
  {\bibfnamefont {Xavier}\ \bibnamefont {Roy}},\ }\bibfield  {title} {\enquote
  {\bibinfo {title} {A few-layer covalent network of fullerenes},}\ }\href
  {\doibase 10.1038/s41586-022-05401-w} {\bibfield  {journal} {\bibinfo
  {journal} {Nature}\ }\textbf {\bibinfo {volume} {613}},\ \bibinfo {pages}
  {71--76} (\bibinfo {year} {2023})}\BibitemShut {NoStop}%
\bibitem [{\citenamefont {Tromer}\ \emph {et~al.}(2022)\citenamefont {Tromer},
  \citenamefont {Ribeiro},\ and\ \citenamefont {Galv{\~a}o}}]{Tromer2022}%
  \BibitemOpen
  \bibfield  {author} {\bibinfo {author} {\bibfnamefont {Raphael~M.}\
  \bibnamefont {Tromer}}, \bibinfo {author} {\bibfnamefont {Luiz~A.}\
  \bibnamefont {Ribeiro}}, \ and\ \bibinfo {author} {\bibfnamefont
  {Douglas~S.}\ \bibnamefont {Galv{\~a}o}},\ }\bibfield  {title} {\enquote
  {\bibinfo {title} {{{A DFT study of the electronic, optical, and mechanical
  properties of a recently synthesized monolayer fullerene network}}},}\ }\href
  {\doibase 10.1016/j.cplett.2022.139925} {\bibfield  {journal} {\bibinfo
  {journal} {Chemical Physics Letters}\ }\textbf {\bibinfo {volume} {804}},\
  \bibinfo {pages} {139925} (\bibinfo {year} {2022})}\BibitemShut {NoStop}%
\bibitem [{\citenamefont {Ribeiro}\ \emph {et~al.}(2022)\citenamefont
  {Ribeiro}, \citenamefont {Pereira}, \citenamefont {Giozza}, \citenamefont
  {Tromer},\ and\ \citenamefont {Galv{\~a}o}}]{Ribeiro2022}%
  \BibitemOpen
  \bibfield  {author} {\bibinfo {author} {\bibfnamefont {L.A.}\ \bibnamefont
  {Ribeiro}}, \bibinfo {author} {\bibfnamefont {M.L.}\ \bibnamefont {Pereira}},
  \bibinfo {author} {\bibfnamefont {W.F.}\ \bibnamefont {Giozza}}, \bibinfo
  {author} {\bibfnamefont {R.M.}\ \bibnamefont {Tromer}}, \ and\ \bibinfo
  {author} {\bibfnamefont {Douglas~S.}\ \bibnamefont {Galv{\~a}o}},\ }\bibfield
   {title} {\enquote {\bibinfo {title} {Thermal stability and fracture patterns
  of a recently synthesized monolayer fullerene network: A reactive molecular
  dynamics study},}\ }\href {\doibase 10.1016/j.cplett.2022.140075} {\bibfield
  {journal} {\bibinfo  {journal} {Chemical Physics Letters}\ }\textbf {\bibinfo
  {volume} {807}},\ \bibinfo {pages} {140075} (\bibinfo {year}
  {2022})}\BibitemShut {NoStop}%
\bibitem [{\citenamefont {Zhao}\ \emph {et~al.}(2024)\citenamefont {Zhao},
  \citenamefont {Zhang}, \citenamefont {Zhao}, \citenamefont {Hu},\ and\
  \citenamefont {Li}}]{Zhao2024a}%
  \BibitemOpen
  \bibfield  {author} {\bibinfo {author} {\bibfnamefont {Xiao-Kun}\
  \bibnamefont {Zhao}}, \bibinfo {author} {\bibfnamefont {Yang-Yang}\
  \bibnamefont {Zhang}}, \bibinfo {author} {\bibfnamefont {Jing}\ \bibnamefont
  {Zhao}}, \bibinfo {author} {\bibfnamefont {Han-Shi}\ \bibnamefont {Hu}}, \
  and\ \bibinfo {author} {\bibfnamefont {Jun}\ \bibnamefont {Li}},\ }\bibfield
  {title} {\enquote {\bibinfo {title} {Understanding the electronic structure
  and chemical bonding in the 2d fullerene monolayer},}\ }\href {\doibase
  10.1021/acs.inorgchem.4c00811} {\bibfield  {journal} {\bibinfo  {journal}
  {Inorg. Chem.}\ }\textbf {\bibinfo {volume} {63}},\ \bibinfo {pages}
  {11572--11582} (\bibinfo {year} {2024})}\BibitemShut {NoStop}%
\bibitem [{\citenamefont {Nagel}\ \emph {et~al.}(1999)\citenamefont {Nagel},
  \citenamefont {Pasler}, \citenamefont {Lebedkin}, \citenamefont {Soldatov},
  \citenamefont {Meingast}, \citenamefont {Sundqvist}, \citenamefont {Persson},
  \citenamefont {Tanaka}, \citenamefont {Komatsu}, \citenamefont {Buga},\ and\
  \citenamefont {Inaba}}]{Nagel1999}%
  \BibitemOpen
  \bibfield  {author} {\bibinfo {author} {\bibfnamefont {P.}~\bibnamefont
  {Nagel}}, \bibinfo {author} {\bibfnamefont {V.}~\bibnamefont {Pasler}},
  \bibinfo {author} {\bibfnamefont {S.}~\bibnamefont {Lebedkin}}, \bibinfo
  {author} {\bibfnamefont {A.}~\bibnamefont {Soldatov}}, \bibinfo {author}
  {\bibfnamefont {C.}~\bibnamefont {Meingast}}, \bibinfo {author}
  {\bibfnamefont {B.}~\bibnamefont {Sundqvist}}, \bibinfo {author}
  {\bibfnamefont {P.-A.}\ \bibnamefont {Persson}}, \bibinfo {author}
  {\bibfnamefont {T.}~\bibnamefont {Tanaka}}, \bibinfo {author} {\bibfnamefont
  {K.}~\bibnamefont {Komatsu}}, \bibinfo {author} {\bibfnamefont
  {S.}~\bibnamefont {Buga}}, \ and\ \bibinfo {author} {\bibfnamefont
  {A.}~\bibnamefont {Inaba}},\ }\bibfield  {title} {\enquote {\bibinfo {title}
  {C$_{60}$ one- and two-dimensional polymers, dimers, and hard fullerite:
  Thermal expansion, anharmonicity, and kinetics of depolymerization},}\ }\href
  {\doibase 10.1103/PhysRevB.60.16920} {\bibfield  {journal} {\bibinfo
  {journal} {Phys. Rev. B}\ }\textbf {\bibinfo {volume} {60}},\ \bibinfo
  {pages} {16920--16927} (\bibinfo {year} {1999})}\BibitemShut {NoStop}%
\bibitem [{\citenamefont {Arvanitidis}\ \emph {et~al.}(2003)\citenamefont
  {Arvanitidis}, \citenamefont {Papagelis}, \citenamefont {Margadonna},
  \citenamefont {Prassides},\ and\ \citenamefont {Fitch}}]{Arvanitidis2003}%
  \BibitemOpen
  \bibfield  {author} {\bibinfo {author} {\bibfnamefont {J.}~\bibnamefont
  {Arvanitidis}}, \bibinfo {author} {\bibfnamefont {Konstantinos}\ \bibnamefont
  {Papagelis}}, \bibinfo {author} {\bibfnamefont {Serena}\ \bibnamefont
  {Margadonna}}, \bibinfo {author} {\bibfnamefont {Kosmas}\ \bibnamefont
  {Prassides}}, \ and\ \bibinfo {author} {\bibfnamefont {Andrew~N.}\
  \bibnamefont {Fitch}},\ }\bibfield  {title} {\enquote {\bibinfo {title}
  {Temperature-induced valence transition and associated lattice collapse in
  samarium fulleride},}\ }\href {\doibase 10.1038/nature01994} {\bibfield
  {journal} {\bibinfo  {journal} {Nature}\ }\textbf {\bibinfo {volume} {425}},\
  \bibinfo {pages} {599--602} (\bibinfo {year} {2003})}\BibitemShut {NoStop}%
\bibitem [{\citenamefont {Kwon}\ \emph {et~al.}(2004)\citenamefont {Kwon},
  \citenamefont {Berber},\ and\ \citenamefont {Tom\'anek}}]{Kwon2004}%
  \BibitemOpen
  \bibfield  {author} {\bibinfo {author} {\bibfnamefont {Young-Kyun}\
  \bibnamefont {Kwon}}, \bibinfo {author} {\bibfnamefont {Savas}\ \bibnamefont
  {Berber}}, \ and\ \bibinfo {author} {\bibfnamefont {David}\ \bibnamefont
  {Tom\'anek}},\ }\bibfield  {title} {\enquote {\bibinfo {title} {Thermal
  contraction of carbon fullerenes and nanotubes},}\ }\href {\doibase
  10.1103/PhysRevLett.92.015901} {\bibfield  {journal} {\bibinfo  {journal}
  {Phys. Rev. Lett.}\ }\textbf {\bibinfo {volume} {92}},\ \bibinfo {pages}
  {015901} (\bibinfo {year} {2004})}\BibitemShut {NoStop}%
\bibitem [{\citenamefont {Brown}\ \emph {et~al.}(2006)\citenamefont {Brown},
  \citenamefont {Cao}, \citenamefont {Musfeldt}, \citenamefont {Dragoe},
  \citenamefont {Cimpoesu}, \citenamefont {Ito}, \citenamefont {Takagi},\ and\
  \citenamefont {Cross}}]{Brown2006}%
  \BibitemOpen
  \bibfield  {author} {\bibinfo {author} {\bibfnamefont {S.}~\bibnamefont
  {Brown}}, \bibinfo {author} {\bibfnamefont {J.}~\bibnamefont {Cao}}, \bibinfo
  {author} {\bibfnamefont {J.~L.}\ \bibnamefont {Musfeldt}}, \bibinfo {author}
  {\bibfnamefont {N.}~\bibnamefont {Dragoe}}, \bibinfo {author} {\bibfnamefont
  {F.}~\bibnamefont {Cimpoesu}}, \bibinfo {author} {\bibfnamefont
  {S.}~\bibnamefont {Ito}}, \bibinfo {author} {\bibfnamefont {H.}~\bibnamefont
  {Takagi}}, \ and\ \bibinfo {author} {\bibfnamefont {R.~J.}\ \bibnamefont
  {Cross}},\ }\bibfield  {title} {\enquote {\bibinfo {title} {Search for
  microscopic evidence for molecular level negative thermal expansion in
  fullerenes},}\ }\href {\doibase 10.1103/PhysRevB.73.125446} {\bibfield
  {journal} {\bibinfo  {journal} {Phys. Rev. B}\ }\textbf {\bibinfo {volume}
  {73}},\ \bibinfo {pages} {125446} (\bibinfo {year} {2006})}\BibitemShut
  {NoStop}%
\bibitem [{\citenamefont {Huang}\ \emph
  {et~al.}(2016{\natexlab{b}})\citenamefont {Huang}, \citenamefont {Lu},
  \citenamefont {Tennessen},\ and\ \citenamefont {M.Rondinelli}}]{Huang2016}%
  \BibitemOpen
  \bibfield  {author} {\bibinfo {author} {\bibfnamefont {Liang-Feng}\
  \bibnamefont {Huang}}, \bibinfo {author} {\bibfnamefont {Xue-Zeng}\
  \bibnamefont {Lu}}, \bibinfo {author} {\bibfnamefont {Emrys}\ \bibnamefont
  {Tennessen}}, \ and\ \bibinfo {author} {\bibfnamefont {James}\ \bibnamefont
  {M.Rondinelli}},\ }\bibfield  {title} {\enquote {\bibinfo {title} {An
  efficient ab-initio quasiharmonic approach for the thermodynamics of
  solids},}\ }\href {\doibase 10.1016/j.commatsci.2016.04.012} {\bibfield
  {journal} {\bibinfo  {journal} {Computational Materials Science}\ }\textbf
  {\bibinfo {volume} {120}},\ \bibinfo {pages} {84--93} (\bibinfo {year}
  {2016}{\natexlab{b}})}\BibitemShut {NoStop}%
\bibitem [{\citenamefont {Pallikara}\ and\ \citenamefont
  {Skelton}(2021)}]{Pallikara2021}%
  \BibitemOpen
  \bibfield  {author} {\bibinfo {author} {\bibfnamefont {Ioanna}\ \bibnamefont
  {Pallikara}}\ and\ \bibinfo {author} {\bibfnamefont {Jonathan~M.}\
  \bibnamefont {Skelton}},\ }\bibfield  {title} {\enquote {\bibinfo {title}
  {{{Phase stability of the tin monochalcogenides SnS and SnSe: a
  quasi-harmonic lattice-dynamics study}}},}\ }\href {\doibase
  10.1039/D1CP02597J} {\bibfield  {journal} {\bibinfo  {journal} {Phys. Chem.
  Chem. Phys.}\ }\textbf {\bibinfo {volume} {23}},\ \bibinfo {pages}
  {19219--19236} (\bibinfo {year} {2021})}\BibitemShut {NoStop}%
\bibitem [{\citenamefont {Gonze}(1995)}]{Gonze1995a}%
  \BibitemOpen
  \bibfield  {author} {\bibinfo {author} {\bibfnamefont {Xavier}\ \bibnamefont
  {Gonze}},\ }\bibfield  {title} {\enquote {\bibinfo {title} {Adiabatic
  density-functional perturbation theory},}\ }\href {\doibase
  10.1103/PhysRevA.52.1096} {\bibfield  {journal} {\bibinfo  {journal} {Phys.
  Rev. A}\ }\textbf {\bibinfo {volume} {52}},\ \bibinfo {pages} {1096--1114}
  (\bibinfo {year} {1995})}\BibitemShut {NoStop}%
\bibitem [{\citenamefont {Baroni}\ \emph {et~al.}(2001)\citenamefont {Baroni},
  \citenamefont {de~Gironcoli}, \citenamefont {Dal~Corso},\ and\ \citenamefont
  {Giannozzi}}]{DFPT}%
  \BibitemOpen
  \bibfield  {author} {\bibinfo {author} {\bibfnamefont {Stefano}\ \bibnamefont
  {Baroni}}, \bibinfo {author} {\bibfnamefont {Stefano}\ \bibnamefont
  {de~Gironcoli}}, \bibinfo {author} {\bibfnamefont {Andrea}\ \bibnamefont
  {Dal~Corso}}, \ and\ \bibinfo {author} {\bibfnamefont {Paolo}\ \bibnamefont
  {Giannozzi}},\ }\bibfield  {title} {\enquote {\bibinfo {title} {Phonons and
  related crystal properties from density-functional perturbation theory},}\
  }\href {\doibase 10.1103/RevModPhys.73.515} {\bibfield  {journal} {\bibinfo
  {journal} {Rev. Mod. Phys.}\ }\textbf {\bibinfo {volume} {73}},\ \bibinfo
  {pages} {515--562} (\bibinfo {year} {2001})}\BibitemShut {NoStop}%
\bibitem [{\citenamefont {Kresse}\ and\ \citenamefont
  {Furthm\"uller}(1996{\natexlab{a}})}]{Kresse1996}%
  \BibitemOpen
  \bibfield  {author} {\bibinfo {author} {\bibfnamefont {G.}~\bibnamefont
  {Kresse}}\ and\ \bibinfo {author} {\bibfnamefont {J.}~\bibnamefont
  {Furthm\"uller}},\ }\bibfield  {title} {\enquote {\bibinfo {title}
  {{Efficient iterative schemes for \textit{ab initio} total-energy
  calculations using a plane-wave basis set}},}\ }\href {\doibase
  10.1103/PhysRevB.54.11169} {\bibfield  {journal} {\bibinfo  {journal} {Phys.
  Rev. B}\ }\textbf {\bibinfo {volume} {54}},\ \bibinfo {pages} {11169--11186}
  (\bibinfo {year} {1996}{\natexlab{a}})}\BibitemShut {NoStop}%
\bibitem [{\citenamefont {Kresse}\ and\ \citenamefont
  {Furthm\"uller}(1996{\natexlab{b}})}]{Kresse1996a}%
  \BibitemOpen
  \bibfield  {author} {\bibinfo {author} {\bibfnamefont {G.}~\bibnamefont
  {Kresse}}\ and\ \bibinfo {author} {\bibfnamefont {J.}~\bibnamefont
  {Furthm\"uller}},\ }\bibfield  {title} {\enquote {\bibinfo {title}
  {Efficiency of ab-initio total energy calculations for metals and
  semiconductors using a plane-wave basis set},}\ }\href {\doibase
  http://dx.doi.org/10.1016/0927-0256(96)00008-0} {\bibfield  {journal}
  {\bibinfo  {journal} {Computational Materials Science}\ }\textbf {\bibinfo
  {volume} {6}},\ \bibinfo {pages} {15 -- 50} (\bibinfo {year}
  {1996}{\natexlab{b}})}\BibitemShut {NoStop}%
\bibitem [{\citenamefont {Dove}(1993)}]{Dove1993}%
  \BibitemOpen
  \bibfield  {author} {\bibinfo {author} {\bibfnamefont {Martin~T.}\
  \bibnamefont {Dove}},\ }\href@noop {} {\emph {\bibinfo {title} {Introduction
  to Lattice Dynamics}}}\ (\bibinfo  {publisher} {Cambridge University Press},\
  \bibinfo {year} {1993})\BibitemShut {NoStop}%
\bibitem [{\citenamefont {Togo}\ \emph {et~al.}(2008)\citenamefont {Togo},
  \citenamefont {Oba},\ and\ \citenamefont {Tanaka}}]{Togo2008}%
  \BibitemOpen
  \bibfield  {author} {\bibinfo {author} {\bibfnamefont {Atsushi}\ \bibnamefont
  {Togo}}, \bibinfo {author} {\bibfnamefont {Fumiyasu}\ \bibnamefont {Oba}}, \
  and\ \bibinfo {author} {\bibfnamefont {Isao}\ \bibnamefont {Tanaka}},\
  }\bibfield  {title} {\enquote {\bibinfo {title} {{First-principles
  calculations of the ferroelastic transition between rutile-type and
  CaCl$_{2}$-type SiO$_{2}$ at high pressures}},}\ }\href {\doibase
  10.1103/PhysRevB.78.134106} {\bibfield  {journal} {\bibinfo  {journal} {Phys.
  Rev. B}\ }\textbf {\bibinfo {volume} {78}},\ \bibinfo {pages} {134106}
  (\bibinfo {year} {2008})}\BibitemShut {NoStop}%
\bibitem [{\citenamefont {Togo}\ and\ \citenamefont {Tanaka}(2015)}]{Togo2015}%
  \BibitemOpen
  \bibfield  {author} {\bibinfo {author} {\bibfnamefont {Atsushi}\ \bibnamefont
  {Togo}}\ and\ \bibinfo {author} {\bibfnamefont {Isao}\ \bibnamefont
  {Tanaka}},\ }\bibfield  {title} {\enquote {\bibinfo {title} {{First
  principles phonon calculations in materials science}},}\ }\href {\doibase
  http://dx.doi.org/10.1016/j.scriptamat.2015.07.021} {\bibfield  {journal}
  {\bibinfo  {journal} {Scripta Materialia}\ }\textbf {\bibinfo {volume}
  {108}},\ \bibinfo {pages} {1--5} (\bibinfo {year} {2015})}\BibitemShut
  {NoStop}%
\bibitem [{\citenamefont {Peng}\ \emph {et~al.}(2019)\citenamefont {Peng},
  \citenamefont {Bravi\ifmmode~\acute{c}\else \'{c}\fi{}}, \citenamefont
  {MacManus-Driscoll},\ and\ \citenamefont {Monserrat}}]{Peng2019}%
  \BibitemOpen
  \bibfield  {author} {\bibinfo {author} {\bibfnamefont {Bo}~\bibnamefont
  {Peng}}, \bibinfo {author} {\bibfnamefont {Ivona}\ \bibnamefont
  {Bravi\ifmmode~\acute{c}\else \'{c}\fi{}}}, \bibinfo {author} {\bibfnamefont
  {Judith~L.}\ \bibnamefont {MacManus-Driscoll}}, \ and\ \bibinfo {author}
  {\bibfnamefont {Bartomeu}\ \bibnamefont {Monserrat}},\ }\bibfield  {title}
  {\enquote {\bibinfo {title} {{{Topological semimetallic phase in
  ${\mathrm{PbO}}_{2}$ promoted by temperature}}},}\ }\href {\doibase
  10.1103/PhysRevB.100.161101} {\bibfield  {journal} {\bibinfo  {journal}
  {Phys. Rev. B}\ }\textbf {\bibinfo {volume} {100}},\ \bibinfo {pages}
  {161101} (\bibinfo {year} {2019})}\BibitemShut {NoStop}%
\bibitem [{\citenamefont {Mortazavi}(2023)}]{Mortazavi2023}%
  \BibitemOpen
  \bibfield  {author} {\bibinfo {author} {\bibfnamefont {Bohayra}\ \bibnamefont
  {Mortazavi}},\ }\bibfield  {title} {\enquote {\bibinfo {title} {Structural,
  electronic, thermal and mechanical properties of c$_{60}$-based fullerene
  two-dimensional networks explored by first-principles and machine
  learning},}\ }\href {\doibase 10.1016/j.carbon.2023.118293} {\bibfield
  {journal} {\bibinfo  {journal} {Carbon}\ }\textbf {\bibinfo {volume} {213}},\
  \bibinfo {pages} {118293} (\bibinfo {year} {2023})}\BibitemShut {NoStop}%
\bibitem [{\citenamefont {Kah}\ \emph {et~al.}(2015)\citenamefont {Kah},
  \citenamefont {Yu}, \citenamefont {Tandy}, \citenamefont {Jayanthi},\ and\
  \citenamefont {Wu}}]{Kah2015}%
  \BibitemOpen
  \bibfield  {author} {\bibinfo {author} {\bibfnamefont {C~B}\ \bibnamefont
  {Kah}}, \bibinfo {author} {\bibfnamefont {M}~\bibnamefont {Yu}}, \bibinfo
  {author} {\bibfnamefont {P}~\bibnamefont {Tandy}}, \bibinfo {author}
  {\bibfnamefont {C~S}\ \bibnamefont {Jayanthi}}, \ and\ \bibinfo {author}
  {\bibfnamefont {S~Y}\ \bibnamefont {Wu}},\ }\bibfield  {title} {\enquote
  {\bibinfo {title} {Low-dimensional boron structures based on icosahedron
  b12},}\ }\href {\doibase 10.1088/0957-4484/26/40/405701} {\bibfield
  {journal} {\bibinfo  {journal} {Nanotechnology}\ }\textbf {\bibinfo {volume}
  {26}},\ \bibinfo {pages} {405701} (\bibinfo {year} {2015})}\BibitemShut
  {NoStop}%
\bibitem [{\citenamefont {Sun}\ \emph {et~al.}(2017)\citenamefont {Sun},
  \citenamefont {Liu}, \citenamefont {Yin}, \citenamefont {Yu}, \citenamefont
  {Li}, \citenamefont {Hang}, \citenamefont {Zhou}, \citenamefont {Yu},
  \citenamefont {Li}, \citenamefont {Tai},\ and\ \citenamefont
  {Guo}}]{Sun2017a}%
  \BibitemOpen
  \bibfield  {author} {\bibinfo {author} {\bibfnamefont {Xu}~\bibnamefont
  {Sun}}, \bibinfo {author} {\bibfnamefont {Xiaofei}\ \bibnamefont {Liu}},
  \bibinfo {author} {\bibfnamefont {Jun}\ \bibnamefont {Yin}}, \bibinfo
  {author} {\bibfnamefont {Jin}\ \bibnamefont {Yu}}, \bibinfo {author}
  {\bibfnamefont {Yao}\ \bibnamefont {Li}}, \bibinfo {author} {\bibfnamefont
  {Yang}\ \bibnamefont {Hang}}, \bibinfo {author} {\bibfnamefont {Xiaocheng}\
  \bibnamefont {Zhou}}, \bibinfo {author} {\bibfnamefont {Maolin}\ \bibnamefont
  {Yu}}, \bibinfo {author} {\bibfnamefont {Jidong}\ \bibnamefont {Li}},
  \bibinfo {author} {\bibfnamefont {Guoan}\ \bibnamefont {Tai}}, \ and\
  \bibinfo {author} {\bibfnamefont {Wanlin}\ \bibnamefont {Guo}},\ }\bibfield
  {title} {\enquote {\bibinfo {title} {Two-dimensional boron crystals:
  Structural stability, tunable properties, fabrications and applications},}\
  }\href {\doibase 10.1002/adfm.201603300} {\bibfield  {journal} {\bibinfo
  {journal} {Adv. Funct. Mater.}\ }\textbf {\bibinfo {volume} {27}},\ \bibinfo
  {pages} {1603300} (\bibinfo {year} {2017})}\BibitemShut {NoStop}%
\bibitem [{\citenamefont {Zhang}\ \emph {et~al.}(2017)\citenamefont {Zhang},
  \citenamefont {Penev},\ and\ \citenamefont {Yakobson}}]{Zhang2017c}%
  \BibitemOpen
  \bibfield  {author} {\bibinfo {author} {\bibfnamefont {Zhuhua}\ \bibnamefont
  {Zhang}}, \bibinfo {author} {\bibfnamefont {Evgeni~S.}\ \bibnamefont
  {Penev}}, \ and\ \bibinfo {author} {\bibfnamefont {Boris~I.}\ \bibnamefont
  {Yakobson}},\ }\bibfield  {title} {\enquote {\bibinfo {title}
  {Two-dimensional boron: structures, properties and applications},}\ }\href
  {\doibase 10.1039/C7CS00261K} {\bibfield  {journal} {\bibinfo  {journal}
  {Chem. Soc. Rev.}\ }\textbf {\bibinfo {volume} {46}},\ \bibinfo {pages}
  {6746--6763} (\bibinfo {year} {2017})}\BibitemShut {NoStop}%
\bibitem [{\citenamefont {Kroto}(1987)}]{Kroto1987}%
  \BibitemOpen
  \bibfield  {author} {\bibinfo {author} {\bibfnamefont {H.~W.}\ \bibnamefont
  {Kroto}},\ }\bibfield  {title} {\enquote {\bibinfo {title} {{{The stability
  of the fullerenes C$_n$, with $n$ = 24, 28, 32, 36, 50, 60 and 70}}},}\
  }\href {\doibase 10.1038/329529a0} {\bibfield  {journal} {\bibinfo  {journal}
  {Nature}\ }\textbf {\bibinfo {volume} {329}},\ \bibinfo {pages} {529--531}
  (\bibinfo {year} {1987})}\BibitemShut {NoStop}%
\bibitem [{\citenamefont {Hallett}\ \emph {et~al.}(1995)\citenamefont
  {Hallett}, \citenamefont {McKay}, \citenamefont {Balm}, \citenamefont
  {Allaf}, \citenamefont {Kroto},\ and\ \citenamefont {Stace}}]{Hallett1995}%
  \BibitemOpen
  \bibfield  {author} {\bibinfo {author} {\bibfnamefont {R.~P.}\ \bibnamefont
  {Hallett}}, \bibinfo {author} {\bibfnamefont {K.~G.}\ \bibnamefont {McKay}},
  \bibinfo {author} {\bibfnamefont {S.~P.}\ \bibnamefont {Balm}}, \bibinfo
  {author} {\bibfnamefont {A.~W.}\ \bibnamefont {Allaf}}, \bibinfo {author}
  {\bibfnamefont {H.~W.}\ \bibnamefont {Kroto}}, \ and\ \bibinfo {author}
  {\bibfnamefont {A.~J.}\ \bibnamefont {Stace}},\ }\bibfield  {title} {\enquote
  {\bibinfo {title} {Reaction studies of carbon clusters},}\ }\href {\doibase
  10.1007/BF01443738} {\bibfield  {journal} {\bibinfo  {journal} {Zeitschrift
  f\"ur Physik D Atoms, Molecules and Clusters}\ }\textbf {\bibinfo {volume}
  {34}},\ \bibinfo {pages} {65--70} (\bibinfo {year} {1995})}\BibitemShut
  {NoStop}%
\bibitem [{\citenamefont {Xu}\ \emph {et~al.}(2023)\citenamefont {Xu},
  \citenamefont {Tian}, \citenamefont {Mu{\~n}oz-Castro}, \citenamefont
  {Frenking},\ and\ \citenamefont {Sun}}]{Xu2023}%
  \BibitemOpen
  \bibfield  {author} {\bibinfo {author} {\bibfnamefont {Yu-He}\ \bibnamefont
  {Xu}}, \bibinfo {author} {\bibfnamefont {Wen-Juan}\ \bibnamefont {Tian}},
  \bibinfo {author} {\bibfnamefont {Alvaro}\ \bibnamefont {Mu{\~n}oz-Castro}},
  \bibinfo {author} {\bibfnamefont {Gernot}\ \bibnamefont {Frenking}}, \ and\
  \bibinfo {author} {\bibfnamefont {Zhong-Ming}\ \bibnamefont {Sun}},\
  }\bibfield  {title} {\enquote {\bibinfo {title} {An all-metal fullerene:
  [k@au$_{12}$sb$_{20}$]$^{5-}$},}\ }\href {\doibase 10.1126/science.adj6491}
  {\bibfield  {journal} {\bibinfo  {journal} {Science}\ }\textbf {\bibinfo
  {volume} {382}},\ \bibinfo {pages} {840--843} (\bibinfo {year}
  {2023})}\BibitemShut {NoStop}%
\bibitem [{\citenamefont {Wu}\ and\ \citenamefont {Peng}(2025)}]{Wu2025}%
  \BibitemOpen
  \bibfield  {author} {\bibinfo {author} {\bibfnamefont {Jiaqi}\ \bibnamefont
  {Wu}}\ and\ \bibinfo {author} {\bibfnamefont {Bo}~\bibnamefont {Peng}},\
  }\bibfield  {title} {\enquote {\bibinfo {title} {Smallest [5,6]fullerene as
  building blocks for 2d networks with superior stability and enhanced
  photocatalytic performance},}\ }\href {\doibase 10.1021/jacs.4c13167}
  {\bibfield  {journal} {\bibinfo  {journal} {J. Am. Chem. Soc.}\ }\textbf
  {\bibinfo {volume} {147}},\ \bibinfo {pages} {1749--1757} (\bibinfo {year}
  {2025})}\BibitemShut {NoStop}%
\bibitem [{\citenamefont {Zhang}\ \emph {et~al.}(2025)\citenamefont {Zhang},
  \citenamefont {Xie}, \citenamefont {Mei}, \citenamefont {Yu}, \citenamefont
  {Li}, \citenamefont {He}, \citenamefont {Fan}, \citenamefont {Zhang},
  \citenamefont {Ricciardulli}, \citenamefont {Samor{\`i}}, \citenamefont
  {Li},\ and\ \citenamefont {Yang}}]{Zhang2025}%
  \BibitemOpen
  \bibfield  {author} {\bibinfo {author} {\bibfnamefont {Yuxuan}\ \bibnamefont
  {Zhang}}, \bibinfo {author} {\bibfnamefont {Yifan}\ \bibnamefont {Xie}},
  \bibinfo {author} {\bibfnamefont {Hao}\ \bibnamefont {Mei}}, \bibinfo
  {author} {\bibfnamefont {Hui}\ \bibnamefont {Yu}}, \bibinfo {author}
  {\bibfnamefont {Minjuan}\ \bibnamefont {Li}}, \bibinfo {author}
  {\bibfnamefont {Zexiang}\ \bibnamefont {He}}, \bibinfo {author}
  {\bibfnamefont {Wentao}\ \bibnamefont {Fan}}, \bibinfo {author}
  {\bibfnamefont {Panpan}\ \bibnamefont {Zhang}}, \bibinfo {author}
  {\bibfnamefont {Antonio~Gaetano}\ \bibnamefont {Ricciardulli}}, \bibinfo
  {author} {\bibfnamefont {Paolo}\ \bibnamefont {Samor{\`i}}}, \bibinfo
  {author} {\bibfnamefont {Mengmeng}\ \bibnamefont {Li}}, \ and\ \bibinfo
  {author} {\bibfnamefont {Sheng}\ \bibnamefont {Yang}},\ }\bibfield  {title}
  {\enquote {\bibinfo {title} {Electrochemical synthesis of 2d polymeric
  fullerene for broadband photodetection},}\ }\href {\doibase
  10.1002/adma.202416741} {\bibfield  {journal} {\bibinfo  {journal} {Adv.
  Mater.}\ }\textbf {\bibinfo {volume} {37}},\ \bibinfo {pages} {2416741}
  (\bibinfo {year} {2025})}\BibitemShut {NoStop}%
\bibitem [{\citenamefont {Wang}\ \emph {et~al.}(2023)\citenamefont {Wang},
  \citenamefont {Zhang}, \citenamefont {Wu}, \citenamefont {Chen},
  \citenamefont {Yang}, \citenamefont {Lu},\ and\ \citenamefont
  {Du}}]{Wang2023}%
  \BibitemOpen
  \bibfield  {author} {\bibinfo {author} {\bibfnamefont {Taotao}\ \bibnamefont
  {Wang}}, \bibinfo {author} {\bibfnamefont {Li}~\bibnamefont {Zhang}},
  \bibinfo {author} {\bibfnamefont {Jinbao}\ \bibnamefont {Wu}}, \bibinfo
  {author} {\bibfnamefont {Muqing}\ \bibnamefont {Chen}}, \bibinfo {author}
  {\bibfnamefont {Shangfeng}\ \bibnamefont {Yang}}, \bibinfo {author}
  {\bibfnamefont {Yalin}\ \bibnamefont {Lu}}, \ and\ \bibinfo {author}
  {\bibfnamefont {Pingwu}\ \bibnamefont {Du}},\ }\bibfield  {title} {\enquote
  {\bibinfo {title} {Few-layer fullerene network for photocatalytic pure water
  splitting into h$_2$ and h$_2$o$_2$},}\ }\href {\doibase
  10.1002/anie.202311352} {\bibfield  {journal} {\bibinfo  {journal} {Angew.
  Chem. Int. Ed.}\ }\textbf {\bibinfo {volume} {62}},\ \bibinfo {pages}
  {e202311352} (\bibinfo {year} {2023})}\BibitemShut {NoStop}%
\bibitem [{\citenamefont {Wang}\ \emph {et~al.}(2024)\citenamefont {Wang},
  \citenamefont {Qiao}, \citenamefont {Zhao}, \citenamefont {Zhang},
  \citenamefont {Zhao}, \citenamefont {Guo}, \citenamefont {Shi}, \citenamefont
  {Liang},\ and\ \citenamefont {Gao}}]{Wang2024b}%
  \BibitemOpen
  \bibfield  {author} {\bibinfo {author} {\bibfnamefont {Qingjie}\ \bibnamefont
  {Wang}}, \bibinfo {author} {\bibfnamefont {Yongqiang}\ \bibnamefont {Qiao}},
  \bibinfo {author} {\bibfnamefont {Kaiyue}\ \bibnamefont {Zhao}}, \bibinfo
  {author} {\bibfnamefont {Peixian}\ \bibnamefont {Zhang}}, \bibinfo {author}
  {\bibfnamefont {Huan}\ \bibnamefont {Zhao}}, \bibinfo {author} {\bibfnamefont
  {Juan}\ \bibnamefont {Guo}}, \bibinfo {author} {\bibfnamefont {Xinwei}\
  \bibnamefont {Shi}}, \bibinfo {author} {\bibfnamefont {Erjun}\ \bibnamefont
  {Liang}}, \ and\ \bibinfo {author} {\bibfnamefont {Qilong}\ \bibnamefont
  {Gao}},\ }\bibfield  {title} {\enquote {\bibinfo {title} {Zero thermal
  expansion in kxmnxin2-x(moo4)3 based materials},}\ }\href {\doibase
  10.1016/j.actamat.2024.120358} {\bibfield  {journal} {\bibinfo  {journal}
  {Acta Materialia}\ }\textbf {\bibinfo {volume} {281}},\ \bibinfo {pages}
  {120358} (\bibinfo {year} {2024})}\BibitemShut {NoStop}%
\bibitem [{\citenamefont {Yao}\ \emph {et~al.}(2019)\citenamefont {Yao},
  \citenamefont {Guan}, \citenamefont {Shiota}, \citenamefont {He},
  \citenamefont {Wang}, \citenamefont {Wu}, \citenamefont {Zheng},
  \citenamefont {Su}, \citenamefont {Yoshizawa}, \citenamefont {Kong},
  \citenamefont {Sato},\ and\ \citenamefont {Tao}}]{Yao2019}%
  \BibitemOpen
  \bibfield  {author} {\bibinfo {author} {\bibfnamefont {Zi-Shuo}\ \bibnamefont
  {Yao}}, \bibinfo {author} {\bibfnamefont {Hanxi}\ \bibnamefont {Guan}},
  \bibinfo {author} {\bibfnamefont {Yoshihito}\ \bibnamefont {Shiota}},
  \bibinfo {author} {\bibfnamefont {Chun-Ting}\ \bibnamefont {He}}, \bibinfo
  {author} {\bibfnamefont {Xiao-Lei}\ \bibnamefont {Wang}}, \bibinfo {author}
  {\bibfnamefont {Shu-Qi}\ \bibnamefont {Wu}}, \bibinfo {author} {\bibfnamefont
  {Xiaoyan}\ \bibnamefont {Zheng}}, \bibinfo {author} {\bibfnamefont
  {Sheng-Qun}\ \bibnamefont {Su}}, \bibinfo {author} {\bibfnamefont {Kazunari}\
  \bibnamefont {Yoshizawa}}, \bibinfo {author} {\bibfnamefont {Xueqian}\
  \bibnamefont {Kong}}, \bibinfo {author} {\bibfnamefont {Osamu}\ \bibnamefont
  {Sato}}, \ and\ \bibinfo {author} {\bibfnamefont {Jun}\ \bibnamefont {Tao}},\
  }\bibfield  {title} {\enquote {\bibinfo {title} {Giant anisotropic thermal
  expansion actuated by thermodynamically assisted reorientation of
  imidazoliums in a single crystal},}\ }\href {\doibase
  10.1038/s41467-019-12833-y} {\bibfield  {journal} {\bibinfo  {journal}
  {Nature Communications}\ }\textbf {\bibinfo {volume} {10}},\ \bibinfo {pages}
  {4805} (\bibinfo {year} {2019})}\BibitemShut {NoStop}%
\bibitem [{\citenamefont {Bond}(2021)}]{Bond2021}%
  \BibitemOpen
  \bibfield  {author} {\bibinfo {author} {\bibfnamefont {Andrew~D.}\
  \bibnamefont {Bond}},\ }\bibfield  {title} {\enquote {\bibinfo {title} {{A
  survey of thermal expansion coefficients for organic molecular crystals in
  the Cambridge Structural Database}},}\ }\href {\doibase
  10.1107/S2052520621003309} {\bibfield  {journal} {\bibinfo  {journal} {Acta
  Crystallographica Section B}\ }\textbf {\bibinfo {volume} {77}},\ \bibinfo
  {pages} {357--364} (\bibinfo {year} {2021})}\BibitemShut {NoStop}%
\bibitem [{\citenamefont {van~der Lee}\ and\ \citenamefont
  {Dumitrescu}(2021)}]{Lee2021}%
  \BibitemOpen
  \bibfield  {author} {\bibinfo {author} {\bibfnamefont {Arie}\ \bibnamefont
  {van~der Lee}}\ and\ \bibinfo {author} {\bibfnamefont {Dan~G.}\ \bibnamefont
  {Dumitrescu}},\ }\bibfield  {title} {\enquote {\bibinfo {title} {Thermal
  expansion properties of organic crystals: a csd study},}\ }\href {\doibase
  10.1039/D1SC01076J} {\bibfield  {journal} {\bibinfo  {journal} {Chem. Sci.}\
  }\textbf {\bibinfo {volume} {12}},\ \bibinfo {pages} {8537--8547} (\bibinfo
  {year} {2021})}\BibitemShut {NoStop}%
\bibitem [{\citenamefont {Saha}\ and\ \citenamefont
  {Veluthaparambath}(2024)}]{Saha2024}%
  \BibitemOpen
  \bibfield  {author} {\bibinfo {author} {\bibfnamefont {Binoy~K.}\
  \bibnamefont {Saha}}\ and\ \bibinfo {author} {\bibfnamefont {Ragima V.~P.}\
  \bibnamefont {Veluthaparambath}},\ }\bibfield  {title} {\enquote {\bibinfo
  {title} {Roles of molecular volume, surface area, heteroatoms, and hydrogen
  bonds on the thermal expansion of organic crystals},}\ }\href {\doibase
  10.1021/acs.cgd.4c00199} {\bibfield  {journal} {\bibinfo  {journal} {Crystal
  Growth \& Design}\ }\textbf {\bibinfo {volume} {24}},\ \bibinfo {pages}
  {3467--3472} (\bibinfo {year} {2024})}\BibitemShut {NoStop}%
\bibitem [{\citenamefont {Blank}\ \emph {et~al.}(1998)\citenamefont {Blank},
  \citenamefont {Buga}, \citenamefont {Dubitsky}, \citenamefont
  {R.~Serebryanaya}, \citenamefont {Popov},\ and\ \citenamefont
  {Sundqvist}}]{Blank1998}%
  \BibitemOpen
  \bibfield  {author} {\bibinfo {author} {\bibfnamefont {V.D.}\ \bibnamefont
  {Blank}}, \bibinfo {author} {\bibfnamefont {S.G.}\ \bibnamefont {Buga}},
  \bibinfo {author} {\bibfnamefont {G.A.}\ \bibnamefont {Dubitsky}}, \bibinfo
  {author} {\bibfnamefont {N.}~\bibnamefont {R.~Serebryanaya}}, \bibinfo
  {author} {\bibfnamefont {M.Yu.}\ \bibnamefont {Popov}}, \ and\ \bibinfo
  {author} {\bibfnamefont {B.}~\bibnamefont {Sundqvist}},\ }\bibfield  {title}
  {\enquote {\bibinfo {title} {High-pressure polymerized phases of c60},}\
  }\href {\doibase 10.1016/S0008-6223(97)00234-0} {\bibfield  {journal}
  {\bibinfo  {journal} {Carbon}\ }\textbf {\bibinfo {volume} {36}},\ \bibinfo
  {pages} {319--343} (\bibinfo {year} {1998})}\BibitemShut {NoStop}%
\bibitem [{\citenamefont {Masago}\ \emph {et~al.}(2006)\citenamefont {Masago},
  \citenamefont {Shirai},\ and\ \citenamefont {Katayama-Yoshida}}]{Masago2006}%
  \BibitemOpen
  \bibfield  {author} {\bibinfo {author} {\bibfnamefont {Akira}\ \bibnamefont
  {Masago}}, \bibinfo {author} {\bibfnamefont {Koun}\ \bibnamefont {Shirai}}, \
  and\ \bibinfo {author} {\bibfnamefont {Hiroshi}\ \bibnamefont
  {Katayama-Yoshida}},\ }\bibfield  {title} {\enquote {\bibinfo {title}
  {Crystal stability of $\ensuremath{\alpha}$- and
  $\ensuremath{\beta}$-boron},}\ }\href {\doibase 10.1103/PhysRevB.73.104102}
  {\bibfield  {journal} {\bibinfo  {journal} {Phys. Rev. B}\ }\textbf {\bibinfo
  {volume} {73}},\ \bibinfo {pages} {104102} (\bibinfo {year}
  {2006})}\BibitemShut {NoStop}%
\end{thebibliography}

%

\clearpage

\appendix

\section{Thermal expansion of 2D icosahedral B$_{12}$ networks}

\begin{figure}[h]
\centering
\includegraphics[width=0.88\linewidth]{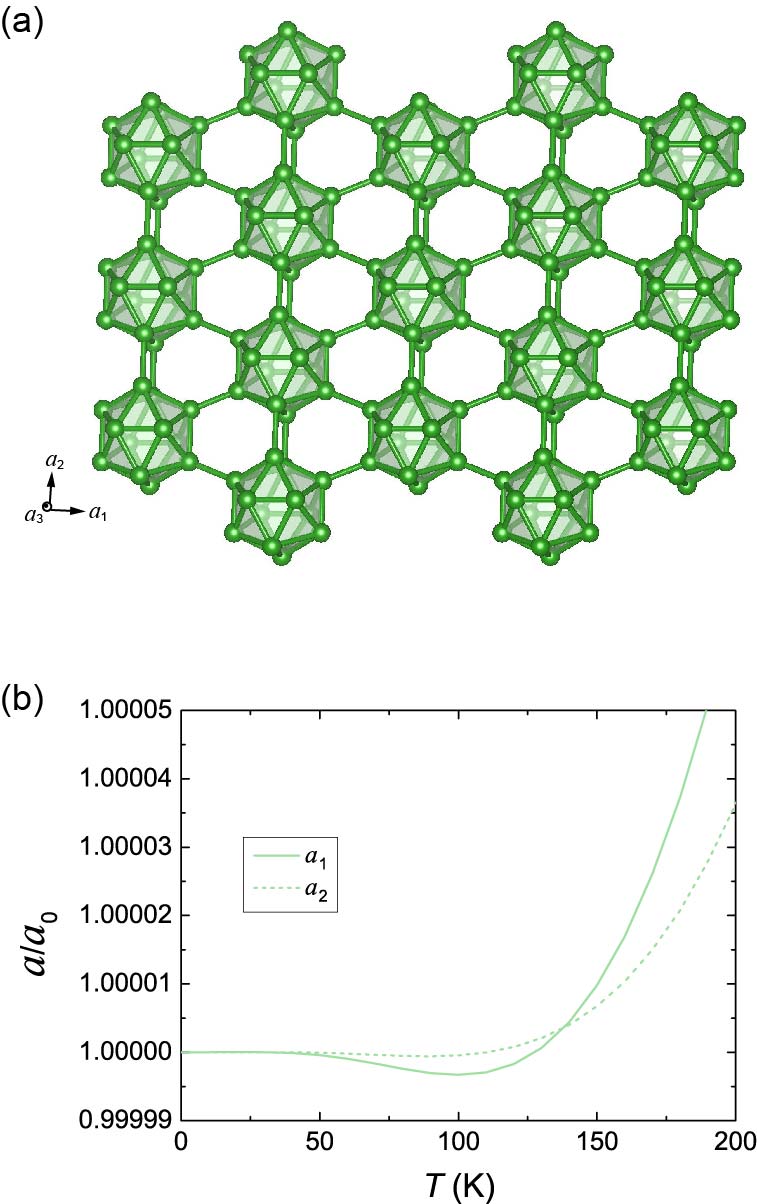}
\caption{
(a) Slightly rotated top view of the crystal structures (to show the double intercluster bonds) and (b) thermal expansion of monolayer icosahedral B$_{12}$ networks.
}
\label{boron} 
\end{figure}

To demonstrate the wide applicability of our design principle beyond carbon-based materials, we provide another example based on the icosahedral B$_{12}$ units in Fig.\,\ref{boron}(a), as these clusters form the most stable allotropes of boron including the $\alpha$ and $\beta$ phases\,\cite{Masago2006}. We explicitly compute thermal expansion of monolayer qHP B$_{12}$ networks and find the same trend in Fig.\,\ref{boron}(b). 

\end{document}